\documentclass[useAMS,usegraphicx, usenatbib]{mn2e}

\usepackage[T1]{fontenc}
\usepackage{ae,aecompl}

\usepackage{aas_macros}
\usepackage{epsfig}
\usepackage{pifont}
\usepackage{verbatim, amsmath, amsfonts, amssymb,amssymb}
\usepackage[utf8x]{inputenc}
\usepackage[usenames,dvipsnames,table]{xcolor}
\usepackage{booktabs, longtable}
\usepackage{multirow}
\usepackage{float}
\usepackage{lscape}
\usepackage{graphicx, subfigure}
\usepackage{rotating}
\usepackage{eufrak}
\usepackage{natbib}
\usepackage{enumerate}
\usepackage{url}
\usepackage{bm}
\usepackage{tikz}
\usepackage{color}
\usepackage[lined]{algorithm2e}
\usepackage{listings}
\usepackage{soul}
\usepackage{tikz}
\usepackage{empheq}
\usepackage[skins,theorems]{tcolorbox}
\usetikzlibrary{positioning, arrows, backgrounds, decorations, 
 				decorations.pathmorphing, decorations.markings}

\tcbset{highlight math style={enhanced,
 colframe=cyan!10!gray,colback=gray!5!white,arc=4pt,boxrule=1pt,
  drop fuzzy shadow}}

\newenvironment{alignteo}%
  {\empheq[box=\tcbhighmath]{align}}
  {\endempheq}

\newcolumntype{a}{>{\columncolor{gray!20}}c}
\newcolumntype{b}{>{\columncolor{white}}c}

\let\en=\ensuremath
\newcommand{\ve}[2]{\en{#1_1},~\en{#1_2},\ \ldots,~\en{#1_{#2}}}
\newcommand{\ma}[3]{\en{#1_{{#2}1}},~\en{#1_{{#2}2},\ \ldots,~\en{#1_{{#2}#3}}}}

\newcommand{\cmark}{\ding{51}}%
\newcommand{\xmark}{\ding{55}}

\def\apj{ApJ}

\newcommand{\xmath}[1]{\ensuremath{#1}\xspace}
\newcommand{\ngc}{\xmath{N_{\mathrm{GC}}}}
\newcommand{\mbh}{\xmath{M_{\mathrm{BH}}}}
\newcommand{\argmin}{\arg\!\min}

\begin{document}
%

\title{The Overlooked Potential of Generalized Linear Models in  Astronomy-III:  Bayesian Negative Binomial Regression and  Globular Cluster Populations}



\pagestyle{myheadings}
\markboth{de Souza et al.}{Bayesian Negative Binomial Regression}

\author[R.S. de Souza]
{\parbox{\textwidth}{\vspace{-.5cm} \Large R.S. de Souza$^{1}$, J.M.  Hilbe$^{2,3}$, B. Buelens$^{4}$,  J.D. Riggs$^{5}$, E. Cameron$^{6}$,        
 E.E.O. Ishida$^{7}$, A.L. Chies-Santos$^{8,9}$,  M. Killedar$^{10}$,
 for the COIN collaboration
}
\vspace{0.3cm}\\
\parbox{\textwidth}{ \small 
$^{1}$MTA E\"otv\"os University, EIRSA ``Lendulet'' Astrophysics Research Group, Budapest 1117, Hungary\\
$^{2}$Arizona State University, 873701,Tempe, AZ 85287-3701, USA\\
$^{3}$Jet Propulsion Laboratory, 4800 Oak Grove Dr., Pasadena, CA 91109, USA\\
$^{4}$Flemish Astronomical Society, 3600 Genk, Belgium\\
$^{5}$Northwestern University, Evanston, IL, 60208, USA\\
$^{6}$Department of Zoology, University of Oxford, Tinbergen Building, South Parks Road, Oxford, OX1 3PS, United Kingdom\\
$^{7}$Max-Planck-Institut f\"ur Astrophysik, Karl-Schwarzschild-Str. 1, 85748 Garching, Germany\\
$^{8}$Departamento de Astronomia, Instituto de F\'isica, Universidade Federal do Rio Grande do Sul, Porto Alegre, R.S, Brazil\\
$^{9}$Instituto de Astronomia, Geof\'isica e Ci\^encias Atmosf\'ericas, Universidade de S\~ao Paulo, S\~ao Paulo, SP, Brazil\\
$^{10}$Universit\"ats-Sternwarte M\"unchen, Scheinerstrasse 1, D-81679, M\"unchen, Germany\\
}
}
\maketitle
\label{firstpage}


\topmargin -1.3cm



\begin{abstract} 
In this paper, the third in a series illustrating the power of generalized linear models (GLMs) for the astronomical community, we elucidate the potential of the class of GLMs which handles count data.  
The size of a galaxy's globular cluster population (\ngc) is a prolonged puzzle  in the astronomical literature. It falls in the category of count data analysis, yet it is usually modelled as if 
it were a continuous response variable.
We have developed a Bayesian negative binomial regression model 
to study the connection between \ngc and the following galaxy  properties: central black hole mass,  dynamical bulge mass, bulge  velocity dispersion, and absolute visual magnitude. 
The methodology introduced herein naturally accounts for  heteroscedasticity, intrinsic scatter, errors in measurements in both axes (either discrete or continuous), and allows modelling  the population of globular clusters on their natural scale as a non-negative integer variable. Prediction intervals of 99 per cent around the trend for expected \ngc comfortably envelope the data, notably including the Milky Way, which has hitherto been considered a problematic outlier. 
Finally, we demonstrate how random intercept models can incorporate  information of each particular galaxy morphological type. 
Bayesian variable selection methodology allows for automatically identifying galaxy types with different productions of GCs, suggesting that on average S0 galaxies have a GC population  35 per cent smaller than other types with similar brightness.

\end{abstract}


\section{Introduction}
\label{sec:intro}

The current era of astronomy marks the transition from a data-deprived field to a data-driven science, for which statistical methods play a central role. 
An efficacious data exploration requires astronomers to go beyond the traditional Gaussian-based models  which are ubiquitous  in  the field. 
Gaussian distributional assumptions  fail to hold when the data to be modelled come from \textit{exponential family} distributions other than the Normal/Gaussian\footnote{The exponential family comprises a set of distributions ranging from both continuous and discrete random variables (e.g., Gaussian, Poisson, Bernoulli, Gamma, etc.)} \citep{Hil12, Hilbe2014}. 
For non-Gaussian regression problems there exist powerful solutions already widely used in medical research \citep[e.g.,][]{lindsey1999}, finance \citep[e.g.,][]{Jong2008}, healthcare \citep[e.g.,][]{gri04} and biostatistics  \citep[e.g.,][]{Marschner12}, but vastly under-utilized to-date in astronomy. These solutions are known as generalized linear models (GLMs). 
Despite the ubiquitous implementation of GLMs in general statistical applications, there have been only a handful of astronomical studies applying  GLM techniques such as logistic regression (e.g.\ \citealt{Rai12,Rai14,lan14,deSouza2014}),  Poisson regression (e.g.\ \citealt{and10}), gamma regression \citep{Elliott2015} and negative binomial (NB) regression \citep{ata14}.  
The methodology discussed herein   focuses  on Bayesian count response models (Poisson and NB), suited to handle discrete, count-based   data sets applied to a catalogue of  globular clusters (GCs). 

Globular clusters are among the oldest stellar 
systems in the Universe (formed at $z>2$, 
\citealt{kruijssen14}), are pervasive in nearby 
massive galaxies, (\citealt{bs06}) and can be found in 
massive galaxy clusters not necessarily associated to 
one of its galaxies (e.g.,  \citealt{durrell14}). 
Hence, understanding their properties is of utmost 
importance for drawing a complete picture of galaxy 
evolution. The past few decades have seen considerable 
interest in the apparent correlation between the mass 
of the black hole at the centre of a galaxy, \mbh, and 
the velocity dispersion of the central stellar bulge, 
$\sigma$ \citep[e.g.,][]{Gebhardt2000}.
As part of the process of understanding the nature and origin of the so-called \mbh--$\sigma$  relation, astronomers have investigated links between other properties of the host galaxy. 
In particular, the correlation between the size of globular cluster populations, \ngc, and \mbh is tight, possibly more so than the \mbh--$\sigma$ relation, and may reflect an underlying connection to the bulge mass, binding energy, host galaxy stellar mass and total luminosity \citep{Burkert2010,Harris2011,Snyder2011,Rhode2012,Harris2013,harris14}. 
This may go some way to explaining the huge range in scales of the regions involved. 
One notorious outlier is our own Milky Way galaxy, for which there are far too many globular clusters given the mass of its central supermassive black hole, despite the fact that both are accurately measured.
Nevertheless, the otherwise small scatter found in such relations deserves a closer look since it cannot be easily explained by simple scaling relation arguments.

The connection between \ngc and the global properties of their  host galaxies is an  extant astronomical puzzle involving count models, but is treated  as a continuous one. Such correlation studies are  commonly based on taking pairs of parameters (x,y) in log-log space and searching for solutions in the normal form $y = \alpha +\beta x$, despite the fact that this regression technique assumes continuous variables and a Gaussian error distribution, e.g.  $\chi^2$--minimisation \citep{Tremaine2002}.

Our method surpasses the previous $\chi^2$-minimisation approach  in several ways. The most obvious being the ability to handle  count  data without the need of logarithmic transformations of a discrete variable. Hence, we can take into account the cases with zero counts, instead of  removing them to accommodate the logarithm transformation, or adding an arbitrary data shift  in the form log(x+$\epsilon$), with  $\epsilon$ commonly taken as unity. Our method naturally handles   errors in variables in both the $x$ and $y$ axes accommodating the heteroscedasticity of the errors in  \ngc\footnote{Heteroscedastic error structures may  remain even after transformation, thus violating the Gaussian assumption of homogeneity of error variance.}. As a further analysis, we introduce one of the most  important extensions of GLMs known as generalized linear mixed models (GLMMs). This is done to include in the model information about each galaxy morphological type, allowing discrimination among classes of objects requiring additional adjustments in their regression coefficients.

The outline of this  paper is  as follows. In section~\ref{sec:glm} we provide a brief introduction of generalized linear models in the context of exponential family distributions. An overview of count data along with Poisson and NB GLMs    are presented in section~\ref{sec:count}. The  dataset  used in our analysis is summarized in section~\ref{sec:catalog}.  In  section~\ref{sec:analysis} we discuss the necessary steps to build our  Bayesian  model. In   section~\ref{sec:GLMMs}, we discuss GLMMs  in the context of  random intercepts models.  Finally in section~\ref{sec:end}, we present our conclusions.

\section{Generalized Linear Models}
\label{sec:glm}

Classical response-with-covariates models, that is, general (not generalized) linear 
models, assume that the response variable and the residual errors, following a normal 
distribution, are linear in the model parameters and have constant variance. This 
allows model parameter estimation with ordinary least squares (OLS) methods. As 
described above, many data sets have response variables that violate one or more of 
these assumptions. While remedial measures such as transformations on the response 
variable or the covariates may be applied, these measures may fall short of 
satisfying the OLS requirements. For data sets for which classical models are ill-suited, the extended class of models, GLMs, are used with model parameters often 
estimated using maximum likelihood methods \citep[for a brief overview of GLMs in an astronomical context, see e.g.,][]{deSouza2014,Elliott2015}.

\citet{nel72} introduced an unification of models characterised by being linear on the systematic component (model predictors). For example logistic and probit analysis for binomial variates, 
contingency tables for multinomial variates, and regression for Poisson- and gamma-
distributed variates, each a form of the GLM. The random response 
variable, $Y_i,~i=1,2,\ldots,n$, may be represented as
\begin{alignteo}
\label{eq:glm}
&Y_i \sim f(\mu_i, a(\phi) V(\mu_i)) , \notag \\
&g(\mu_i) = \eta_i,  \\ 
&\eta_i \equiv \boldsymbol{x}_i^T \boldsymbol{\beta} = \beta_0+\beta_1x_1+\cdots+\beta_px_p. \notag 
\end{alignteo}

In equation~(\ref{eq:glm}), $f$ denotes a response variable distribution from the 
exponential family (EF), $\mu_i$ is the response variable mean, $\phi$ is the EF 
dispersion parameter in the dispersion function $a(\cdot)$, $V(\mu_i)$ is the 
response variable variance function, $\eta_i$ is the linear predictor, the 
$\boldsymbol{x}_i^T=\{\ma x i p \}^T$ is the vector of explanatory variables 
(covariates or predictors), $\boldsymbol{\beta}=\{\ve \beta p \}$ is the vector of covariates 
coefficients, and $g(\cdot)$ is the link function, which connects the mean to the predictor. If $V(\mu_i)$ is a constant 
for all $\mu_i$, then the mean and variance of the response are independent, which 
allows using a Gaussian response variable. If the response is Gaussian, then $g(\mu) = \mu$. The general form of the GLM 
thus allows Gaussian family, $\mathcal{N}$,   linear regression as a subset, taking the  form:
\begin{alignteo}
\label{eq:lm}
&Y_i \sim \mathcal{N}(\mu_i, \sigma^2) , \notag \\
 &\mu_i =  \beta_0+\beta_1x_1+\cdots+\beta_px_p.  
\end{alignteo}
The subset of GLMs for count data are the Poisson regression models and the 
several incarnations of the NB regressions. Poisson regression models assume the 
count  response variable follows a Poisson probability distribution function. 
Similarly, the NB regression models assume the count response variable follows a 
 NB probability distribution function. Descriptions of the Poisson and NB 
models follow.

\begin{table*}
\caption{Main assumptions of each regression model family.}
\label{tab:assumptions}
\begin{tabular}{l | a | b | a |b}
\hline
\rowcolor{cyan!20}
 & Normal & Log-normal & Poisson & Negative binomial   \\
\hline
Response variable & Real & Positive & Non-negative integer & Non-negative 
integer  \\
Null values &\cmark & \xmark & \cmark & \cmark  \\
Sample variance & Homoscedastic & Homoscedastic & Heteroscedastic & 
Heteroscedastic  \\ 
Overdispersion & \xmark & \xmark & \xmark & \cmark  \\\hline
\end{tabular}
\end{table*}

\section{Modelling count data}
\label{sec:count}

Astronomical quantities  can be measured on different scales: nominal (e.g., classes 
of objects: Type Ia/II supernovae, elliptical/spiral galaxies);  ordinal, (e.g., ordering planets according to their size or distance to the  star);   and metric (e.g., galaxy mass, stellar  temperature).  
Observations that have only right-skewed, non-negative integer values belong to a subclass of the 
metric scale  known as count data.  Distances between counts are meaningful, hence 
the  counts are metric,  but they are not continuous and must be  treated as such. Astronomical  count data are often log-transformed to satisfy Gaussian parametric test assumptions rather than modelled on the basis of a count distribution. Despite the fact that GLMs are better suited to describe count data, a log-transformation of counts has the additional problem of dealing with zeros as observations. With just one observation with value zero, the entire data set needs to be shifted by adding an arbitrary value  before transformation. It is well known that such transformations perform poorly, leading to bias in the estimated parameters 
\citep{OHara2010}. 

We begin our discussion of regression models for count data with the subset of 
GLMs known as Poisson regression. A common condition accompanying count data is 
overdisperson, it occurs when the variance exceeds the mean. This condition in Poisson regression suggests that remedial 
measures, such as the use of NB regression, may be appropriate.

\subsection{Poisson Regression}

Poisson regression was the first model specifically used to deal with count data and 
still stands as basis for many types of analyses. It assumes a discrete response 
described by a single parameter distribution which represents the mean or 
rate, $\mu$; i.e., the expected number of times an event occurs within a fixed time-interval.
Another 
important feature is the assumption of equidispersion which implies  the equality 
of mean and variance, and can be quantified by the Pearson $\chi^2$ dispersion 
statistic (see \S~\ref{sec:overdispersion}.). 
The Poisson distribution function is typically displayed as
\begin{equation}
f(y;\mu) = \frac{\mu^y\mathrm{e}^{-\mu}}{y!},
\end{equation}
where the mean and variance are given by
\begin{equation}
\rm{Mean}  = \mu,\qquad \rm{Variance} = \mu,  
\end{equation}
representing a particular case of equation~(\ref{eq:glm}) with  $V(\mu) = \mu$ and 
$a(\phi) = 1$. Thus, a regression equation derived from equation~(\ref{eq:glm})  may be used as a GLM for a count response, $y$. The usual link function, $g(\mu)$, is the natural log function such that $\mu = e^\eta$ \citep[see e.g.,][]{Hil12}.
It is worth noting that GLMs are not simple log  transforms of the response variable, but rather, the expected counts from a Poisson  regression is an exponentiated linear function of $\eta$, thereby keeping the response variable on its original scale.
Often, count data do not enjoy the Poisson assumption of equidispersion resulting in a Poisson dispersion statistic (see section~\ref{sec:overdispersion})  with a value greater than one.

\subsection{Overdispersion}
\label{sec:overdispersion}

Overdispersion in Poisson models occurs when the response variance is
greater than the mean.  It may arise when there are violations in the distributional 
assumptions of the
data such as when the data are clustered, thereby violating the likelihood requirement of the 
independence of observations. Overdispersion may cause standard errors of the 
estimates to be deflated or
underestimated, i.e. a variable may appear to be a significant predictor when it is 
in fact not. A key approach for checking overdispersion is  by means of the 
dispersion statistic, $\mathcal{D}$,
\begin{equation}
\mathcal{D} = \frac{\chi^2}{N-N_p}, 
\end{equation}
where $N$ is the number of observations and $N_p$ is the number of parameters in the model. Then $N-N_p$ represents the residual degrees of freedom.  For a Poisson GLM, the Pearson $\chi^2$ value is
\begin{equation}
\chi^2 = \sum_{i=1}^{N}\frac{(Y_i-\mu_i)^2}{\mu_i}, 
\end{equation}
where $Y_i$ represents  the observed values, and   $\mu_i$ is the  mean  and variance of $Y_i$. 
Poisson overdispersion occurs when the variation in the data exceeds the expected  variability based on the Poisson distribution, resulting in $\mathcal{D}$  being greater than 1.
Small amounts of overdispersion are of little concern; a rule of thumb is: if $\mathcal{D} > 1.25$, then a correction may be warranted \citep{Hilbe2014}.

If overdispersion is observed,  then there are several corrective measures in common practice. Options are  adjusting the standard errors by scaling, applying  sandwich or robust standard errors, or bootstrapping standard errors for the model. However, only the standard
errors will be adjusted and not the  regression coefficients, $\mathbf{\beta}$, which often  can  be affected  by overdispersion as well \citep[e.g.,][]{Hilbe2011}. 
This paper examines the efficacy of using Bayesian estimation methods on a more general discrete  distribution  known as the NB. The NB distribution contains a second parameter called the dispersion or heterogeneity parameter which is used to accommodate Poisson overdispersion as described below.

\subsection{Negative Binomial Regression}

The NB distribution  has long been recognized as a full member of the exponential
family, originally representing the probability of observing $y$ failures before the $rth$
 success in a series of Bernoulli trials. 
It can also be formulated as a Poisson model with gamma heterogeneity \citep{Hilbe2011}.
The NB model, as a Poisson–gamma mixture model, is appropriate
to use when the overdispersion in an otherwise Poisson model is thought
to take the form of a gamma shape or distribution, i.e., $a(\phi) = 1 / k$, with  $k > 0$. 
The NB probability distribution function is then  given by: 
\begin{equation}
f(y;k,\mu) = \frac{\Gamma(y+k)}{\Gamma(k)\Gamma(y+1)}\left(\frac{k}{\mu+k}\right)^k\left(1-\frac{k}{\mu+k}\right)^y.
\end{equation}
The distribution function has two parameters, $\mu$ and $k$, allowing more flexible models than the Poisson distribution. The symbol $\Gamma$ represents the  gamma function\footnote{If $n$ is a positive integer, $\Gamma (n) = (n-1)!$. }.
The mean and variance are given by
\begin{equation}
\rm{Mean}  = \mu;\qquad \rm{Variance} = \mu+\frac{\mu^2}{k} = \mu+\alpha\mu^2.
\end{equation}
The NB distribution has distributional assumptions similar to the Poisson distribution with the exception that it has a dispersion parameter $\alpha = 1/k$ to accommodate wider count  distribution shapes than allowed by the Poisson model.  As the dispersion parameter, $\alpha$,  approaches 0, $\displaystyle \lim_{\alpha \to 0}\alpha\mu^2 = 0$  
or  $\displaystyle \lim_{k \to \infty}\mu^2/k = 0$, then  the variance equals the mean which  recovers the Poisson distribution.

It should be  noted that if different clusters of counts have different gamma shapes, indicating differing degrees of correlation within data, and if the NB Pearson $\chi^2$ dispersion statistic is greater than one,  then the NB model may itself be overdispersed; i.e the data may be both Poisson and NB
overdispersed. Random effects and mixed effects Poisson and NB models are then reasonable alternatives \citep{Hilbe2014}.

An additional situation should also be mentioned. If the Poisson dispersion statistic is less than one, this is evidence of Poisson under-dispersed data. The NB model is not appropriate for handling Poisson under-dispersion; however, the generalized Poisson model is. 
We do not discuss under-dispersed data in this article, but the subject warrants future study as to how it applies to astrophysical data. 
To guide the reader, Table~\ref{tab:assumptions} displays the main assumptions of 
the OLS, OLS with a log-transformed response variable,  Poisson,  and NB regression models discussed in the previous sections.

%
\section{Dataset}
\label{sec:catalog}

As a study case, we use the catalogue of globular clusters  presented in \cite{Harris2013} (see also \citealt{harris14})\footnote{The complete catalogue can be obtained  at \url{http://www.physics.mcmaster.ca/~harris/GCS_table.txt}.}. The data are composed of  422 galaxies with published measurements of their globular cluster  populations. There is a range of galaxy morphologies from which we indexed  247 as  elliptical (E), 94 as lenticular (S0),  55 as spirals (S) and 26 as irregulars (Irr) galaxies for illustrative purposes. Note that the original catalogue presents 69 different subcategories of morphological classifications which will be discussed in section~\ref{sec:GLMMs}. 
This  is a compilation of literature data from a variety of sources obtained with the Hubble Space Telescope as well as a wide range of other ground based facilities. 
Beyond  \ngc, we select the following properties for our analysis:  central
black hole mass,  dynamical bulge mass, bulge  velocity dispersion, and absolute visual magnitude as described  in Table~\ref{tab:var}.

\begin{table}
\centering
\caption{Summary of the  parameters used in this work from the catalogue of globular clusters compiled by \citeauthor{Harris2013}.  }
\label{tab:var}
\begin{tabular}{l | a }
\hline
\rowcolor{cyan!20}
Parameter  & Definition    \\
\hline
\ngc   & Number of globular clusters  \\ 
$M_V$  & Absolute visual magnitude  \\ 
$\sigma$ & Bulge  velocity dispersion   \\
\mbh   & Central black hole mass   \\
$M_{\rm dyn}$ & Dynamical mass   \\
$\epsilon_{\ngc}$ & Uncertainty in \ngc\\
$\epsilon_{M_V}$ & Uncertainty in $M_V$\\
$\epsilon_{\sigma}$ & Uncertainty in $\sigma$\\
$\epsilon_{\mbh}$ & Uncertainty in \mbh\\ 
\hline
\end{tabular}

\end{table}

\section{Modelling the population size of globular clusters}
\label{sec:analysis}

\begin{figure}
\centering
\includegraphics[width=1\columnwidth]{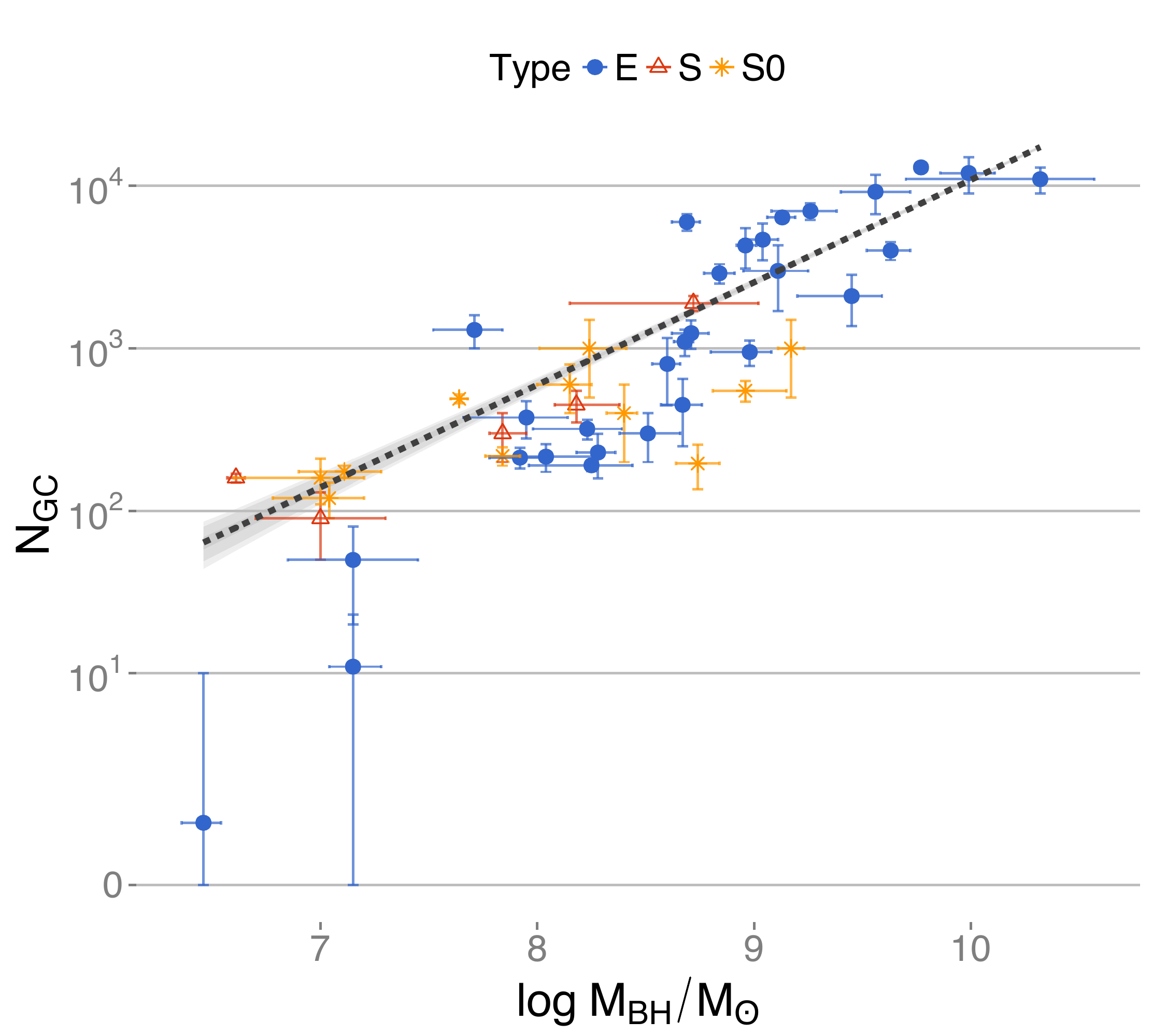}
\caption{Total number of globular clusters,  \ngc, plotted versus the central black hole mass, \mbh. The dashed line represents the expected value of \ngc for each value of \mbh using Poisson GLM regression, 
while the shaded areas depicts 50\%, 95\%,  and 99\% prediction intervals.  Galaxy types are coded by shape and colour as follows: Ellipticals (E; blue solid circles),  spirals (S; red open triangles), and  lenticulars (S0; orange asterisks). An  ArcSinh transformation is applied in the y-axis for better visualization of the whole range of \ngc values,   including the null ones.
}
\label{fig:fitpois}
\end{figure}

\begin{figure}
\centering
\includegraphics[width=0.85\columnwidth]{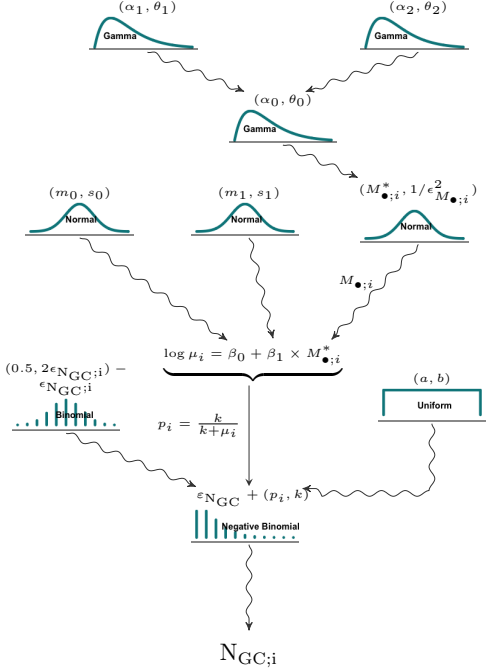}
\caption{A graphical model of equation~(\ref{eq:HBM}) representing the hierarchy of  dependencies for a data set of galaxies indexed  by the subscript $i$. The sinusoidal curves represent stochastic dependencies, while straight arrows a deterministic ones.  To save space, we replace \mbh by $M_{\bullet}$ in the diagram.  
}
\label{fig:HBM}
\end{figure}


Within this section we demonstrate the application of Bayesian GLM regression for modelling the  relationship between \ngc and the following  galaxy properties: \mbh, $\sigma$, $M_V$ and $M_{\rm dyn}$. Hereafter, unless otherwise stated, the analysis is made   using a  sub-sample of 45 objects from which we have observations for all the property predictors. 
In section~\ref{sec:all} an additional  analysis uses the entirety of the available data.

A few common terms in statistical modelling need to be reviewed to facilitate our model applications.
The  analysis  focus is the prediction of  \ngc as a function of the global galaxy properties. Therefore,  \ngc   represents the count (i.e., a non-negative integer) response variable,  while $M_V$, \mbh and $M_{\rm dyn}$ are  interchangeably   called covariates, explanatory variables or predictors. If included in the model, the galaxy morphological type is also considered a nominal categorical predictor  (see  section~\ref{sec:count}). The whole  analysis is  performed using JAGS (Just Another Gibbs Sampler)\footnote{http://CRAN.R-project.org/package=rjags}, a  program  for analysis of Bayesian hierarchical models using a Markov Chain Monte Carlo (MCMC) framework\footnote{Note that count models can be approached by other methods, such as a full maximum likelihood algorithm \citep[see][for a review]{Hilbe2011}}. For each regression case, we initiate three  Markov chains  by starting the Gibbs samples at different initial values sampled from a  normal distribution with zero mean and standard deviation of 10. The initial adapting and burning phases were set to 22,000 steps followed subsequently by 50,000 steps, which was sufficient to guarantee convergence of each chain for all studied cases.

We now use the relationship between \mbh and \ngc as an example to illustrate how the statistical  model is built.   To motivate the use of the more general NB distribution, we start the analysis assuming a GLM  Poisson regression model neglecting the uncertainties  in measurements at this stage for simplicity\footnote{Neglecting the errors at this point does not affect the conclusions regarding the level of Poisson overdispersion.}. 
This  leads to the following model: 
\begin{alignteo}
\label{eq:pois}
  &N_{\mathrm{GC};i}\sim \rm Poisson(\mu_i); \notag   \\
    &\mu_i = e^{\eta_i}; \notag  \\ 
  &\eta_i =  \beta_0+\beta_1\times M_{\mathrm{BH};i};  \\
 &\beta_0 \sim N(0,10^{6});\notag \\
  &\beta_1 \sim N(0,10^{6});\notag \\
  &i = 1,\cdots,N. \notag
\end{alignteo}
This set of equations reads as  follows: each galaxy in the dataset, composed of  $N$ objects,  has its globular cluster population sampled from a Poisson distribution whose expected value,  $\mu$,  relates to the central black hole mass through a linear relation expressed by  $\eta$. Since we don't have  previous information about the values of the  coefficients  $\beta_0$ and $\beta_1$,   we   assigned non-informative  Gaussian priors  with zero mean and standard deviation equal to  $10^6$. We refer the reader to appendix~\ref{app:JAGS} for an example of how to implement a Poisson GLM  in  JAGS. 
The fitted curve for this model is displayed in Fig.~\ref{fig:fitpois}. The grey shaded areas represent 50\%,  95\%,  and  99\% prediction intervals, which are the regions where a future observation will fall with these given  probabilities\footnote{Not to be confused  with the commonly used   confidence interval in frequentist statistics. A 95\% confidence interval will contain the sample  mean with 95\% probability. In other words, a larger number of repeated samples from the data would contain the sample mean 95\% of the time.}. Note that the areas in the plot are too narrow to be visually discriminated.
A visual inspection clearly indicates that the Poisson model isn't adequate to explain the data variability since most of the data fall outside the three prediction intervals. Also,  the dispersion statistic for this model is $\mathcal{D} = 1039$,  which is a strong indication of an inadequate model.  All other covariates, $\sigma$, and $M_V$ and $M_{\rm dyn}$, lead to models with similarly high levels  of Poisson  overdispersion. Hence, hereafter  we discuss construction of the full model based on the NB  family to mitigate overdispersion and to include the uncertainties in the observational quantities. Unlike the Poisson model, by employing a NB distribution we allow the incidence rate of globular clusters to  be itself a random variable. 

Continuing with our working example, we keep the discussion using the relationship between \ngc and \mbh, but see appendix~\ref{app:hbm} for  descriptions of the other models. The first step is to understand how to include information about the uncertainties in the measurements \citep[see e.g.,][ for a review of measurement errors in astronomy]{Andreon2013}.  Measurement errors  in the response count variable are the trickiest part to be modelled. The  classical  model  with an additive error term $y = y^{*} \pm \varepsilon$ is inappropriate since it does not ensure that the observed value 
$y$ is non-negative. 
The appropriate  model  is described below and its graphical representations  are  displayed in Fig.~\ref{fig:HBM}:
\begin{alignteo}
\label{eq:HBM}
  &N_{\mathrm{GC};i}\sim \rm NB(p_i,k); \notag   \\
  &p_i =  \frac{k}{k+\mu_i}; \notag \\
  &\mu_i = e^{\eta_i} + \epsilon_{N_{GC};i}; \notag  \\ 
  &\eta_i =  \beta_0+\beta_1\times M_{\mathrm{BH};i}^*; \notag \\
  & k \sim \mathcal{U}(0,5); \notag \\
  &M_{\mathrm{BH};i} \sim \mathcal{N}(M_{\mathrm{BH};i}^*,e_{\mathrm{BH};i}^2);\notag \\ 
  &\epsilon_{N_{\mathrm{GC}};i} \sim \mathcal{B}(0.5, 2 e_{N_{GC};i}) - e_{N_{\mathrm{GC}};i}; \\  
  &\beta_0 \sim \mathcal{N}(0,10^{6});\notag \\
  &\beta_1 \sim \mathcal{N}(0,10^{6});\notag \\
  &M_{\mathrm{BH};i}^* \sim \Gamma(\alpha_0,\theta_0); \notag \\
  &\alpha_0 \sim \Gamma(0.01,0.01); \notag \\
  &\theta_0 \sim \Gamma(0.01,0.01); \notag \\
  &i = 1,\cdots,N.\notag
\end{alignteo}
The above  is slightly more complex than the model   displayed in equation~(\ref{eq:pois}) and reads as follows.  Each galaxy in the dataset with $N$ objects, has its globular cluster population sampled from a NB distribution whose expected value,  $\mu$,  relates to the central black hole mass through the linear predictor  $\eta$. The additional transformation $p_i = k/(k+\mu_i)$ is required due to  how the NB distribution is parametrized in JAGS.  
The uncertainties related to the counts, $\epsilon_{N_{\mathrm{GC}};i}$, are taken to be associated with the mean, $\mu$,  of the NB  distribution and are modelled using a shifted binomial distribution, $\mathcal{B}$,  with zero mean and taking on integer values in the range $[-e_{N_{GC};i}, +e_{N_{GC};i}]$ \citep[see e.g., Chapter 13 from][ from which this approach is loosely based.]{Cameron2013}. Uncertainties associated with the observed predictor $M_{BH;i}$ are modelled using a Gaussian distribution with unobserved mean given by the ``true black hole mass'',  $M_{BH;i}^*$, and standard deviations given by the reported uncertainties in the observed black hole mass, $e_{M_{BH};i}$. Since $M_{BH;i}^*$ is itself an unobserved variable, we add a non-informative $\Gamma$  prior on top of which  we added  non-informative hyperpriors for the shape, $\alpha_0$,  and rate, $\theta_0$, parameters of the  $\Gamma$  distribution. The choice of a $\Gamma$  prior is motivated by the fact that the black hole mass is a continuous, but non-negative quantity which makes $\Gamma$  a more suitable  distribution. For the shape parameter $k$, we assigned a non-informative uniform prior, $\mathcal{U}$,  as suggested in \citet{zuu13}. 
For the  coefficients  $\beta_0$ and $\beta_1$  we assigned  non-informative  Gaussian priors  with zero mean and standard deviation equal to  $10^6$.

\begin{figure*}
\centering
\includegraphics[width=1\columnwidth]{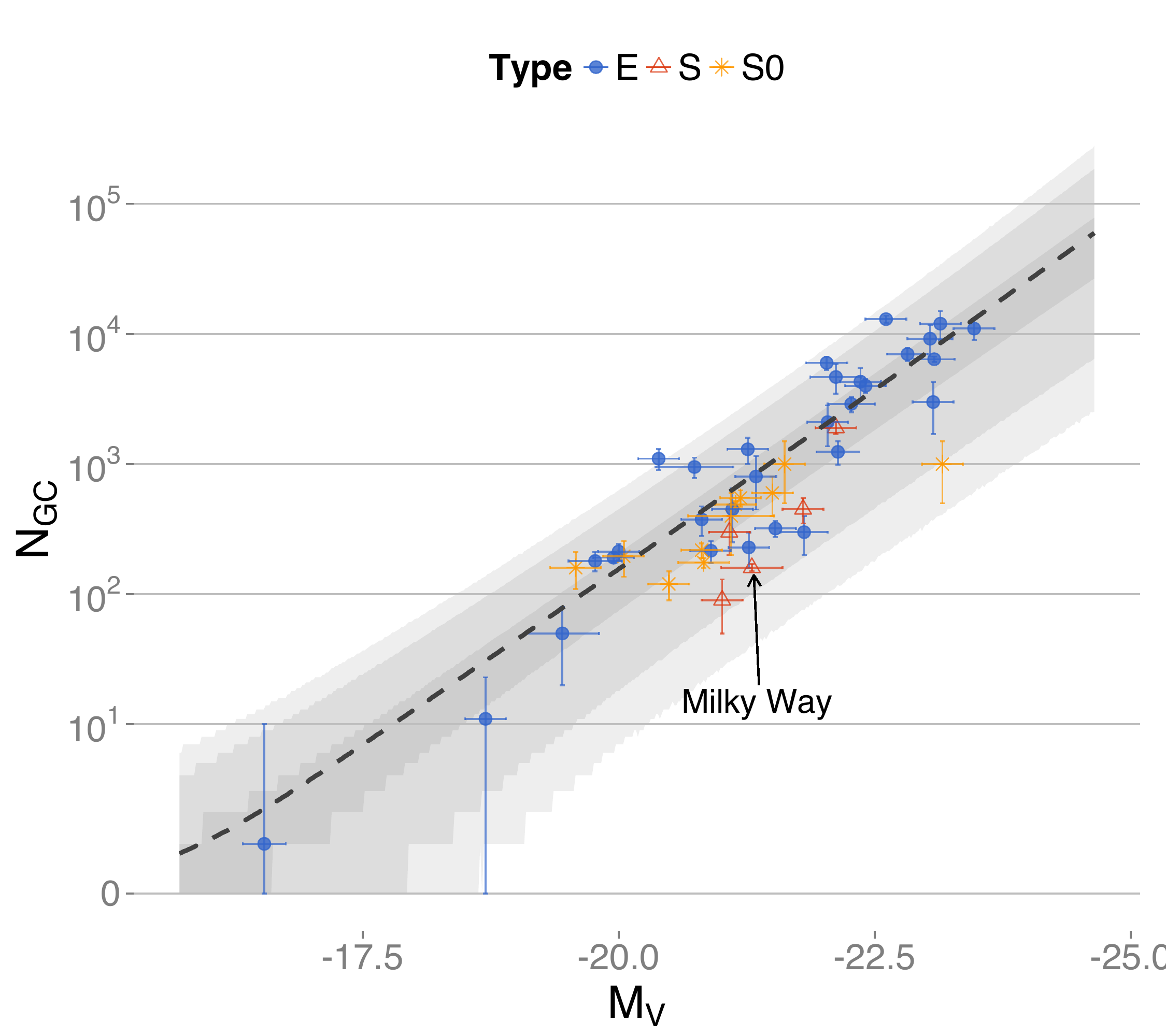}
\includegraphics[width=1\columnwidth]{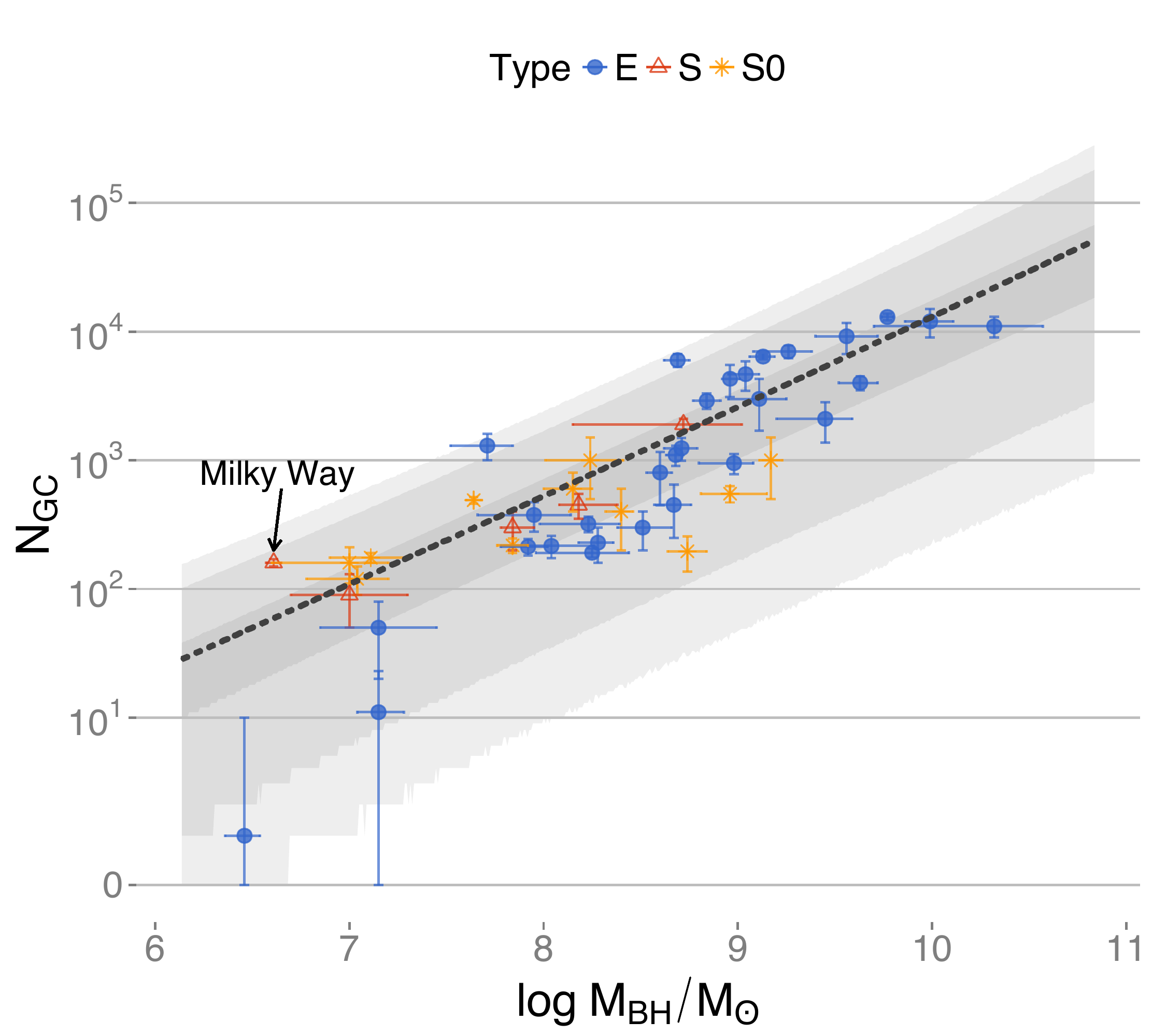}
\includegraphics[width=1\columnwidth]{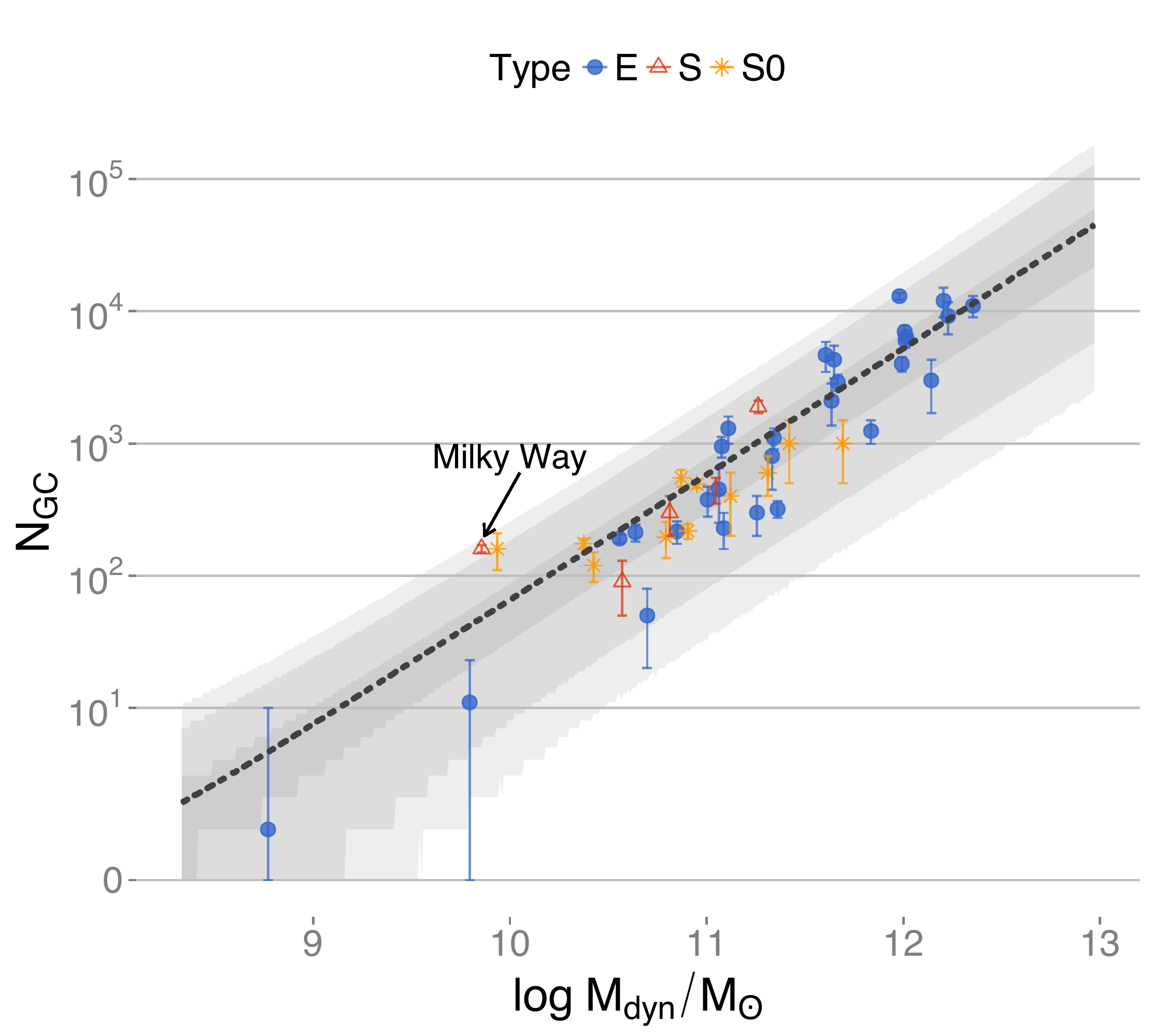}
\includegraphics[width=1\columnwidth]{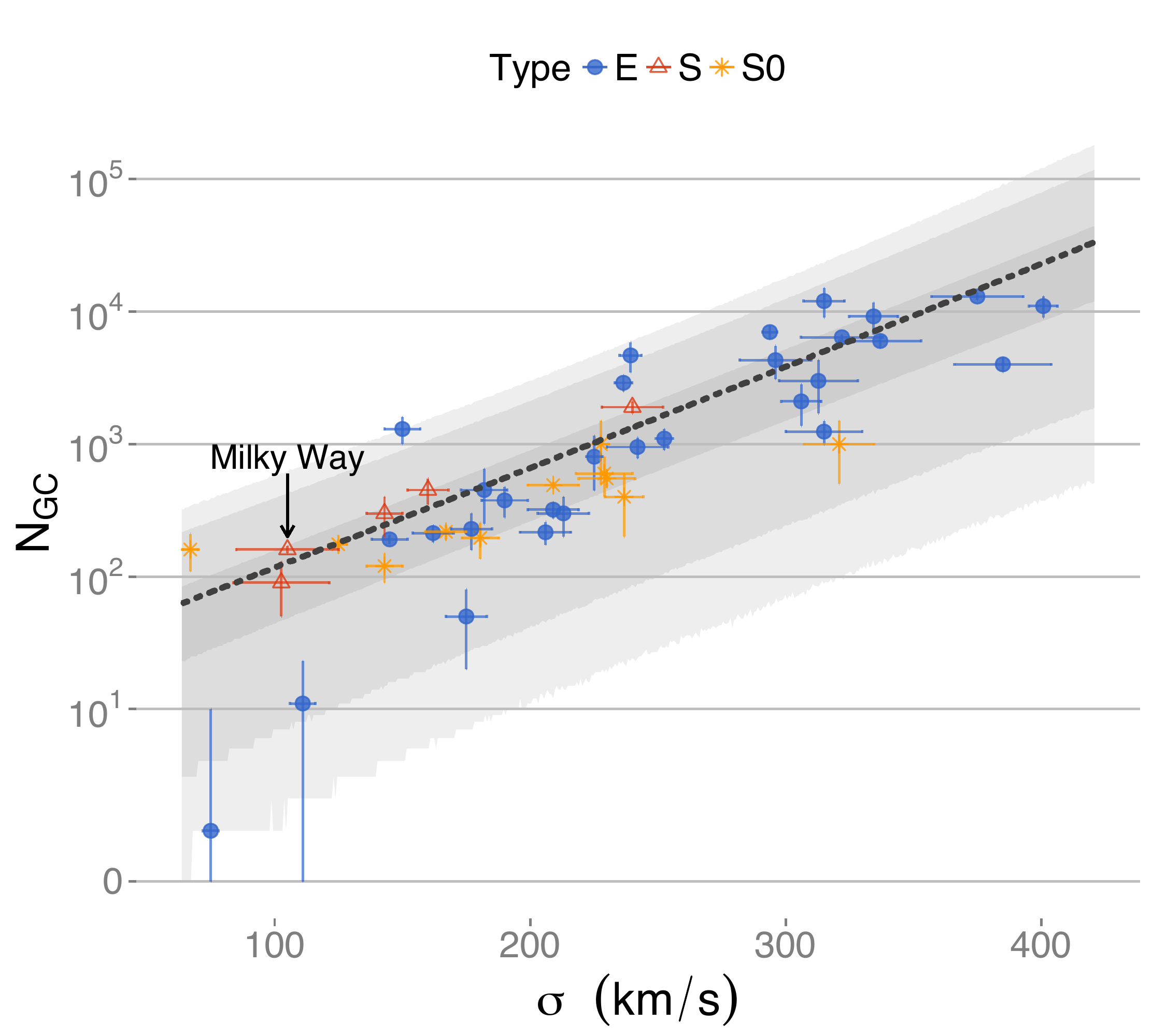}
\caption{Globular cluster population, \ngc plotted against visual absolute magnitude ($M_V$; top left panel), black hole mass (\mbh; top right panel), dynamical mass ($M_{\rm dyn}$; bottom  left panel), and bulge velocity dispersion ($\sigma$; bottom  right panel).   In each panel the dashed line  represents the expected value of \ngc for each value of the covariate using negative binomial GLM regression, 
while the shaded areas depicts 50\%,  95\% , and  99\% prediction intervals. Galaxy types are coded by shape and colour as follows: Ellipticals (E; blue solid circles),  spirals (S; red open triangles), and  lenticulars (S0; orange asterisks). An  ArcSinh transformation is applied in the y-axis for better visualization of the whole   range of \ngc values,   including the null ones.}
\label{fig:fit_negbin}
\end{figure*}

Adapting the model above for each combination of \ngc and a given galaxy property generates the  fitted curves  displayed in Fig.~\ref{fig:fit_negbin}. The grey shaded area represents 
50\%,  95\%, and  99\% prediction intervals, while the dashed line  represents the expected value of \ngc for each value of the covariate. Note the remarkable agreement between the model and the observed values  with prediction intervals  enclosing the entirety of the data, including objects that have been previously declared outliers and even removed from analysis, such as our own Milky Way \citep[e.g.,][]{Burkert2010,Harris2011,harris14}.

\subsection{Fit diagnostics}

\begin{figure}
\centering
\includegraphics[width=1\columnwidth]{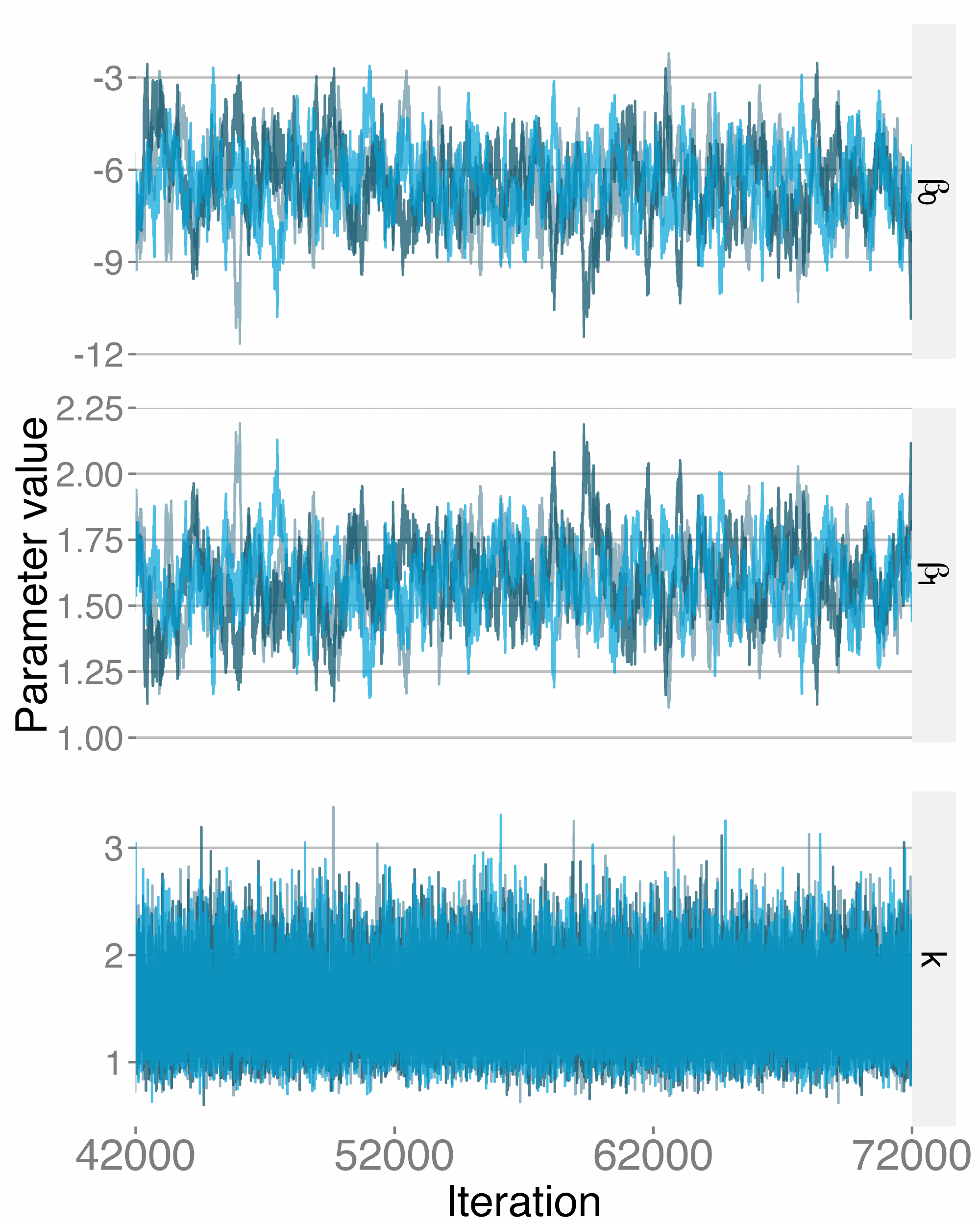}

\caption{Illustration of MCMC diagnostics. Three chains were generated by starting the Gibbs algorithm at different initial values sampled from  a normal distribution with zero mean and standard deviation 10. Steps  
42,000-72,000 are shown here. The figure displays the results for the model \ngc vs $M_{\rm BH}$, with the traceplots  for  $\beta_0$, $\beta_1$ and $k$ displayed from top to bottom. }
\label{fig:traceplot}
\end{figure}

\begin{figure}
\centering
\includegraphics[trim = 5mm 25mm 5mm 25mm, clip,width=1\columnwidth]{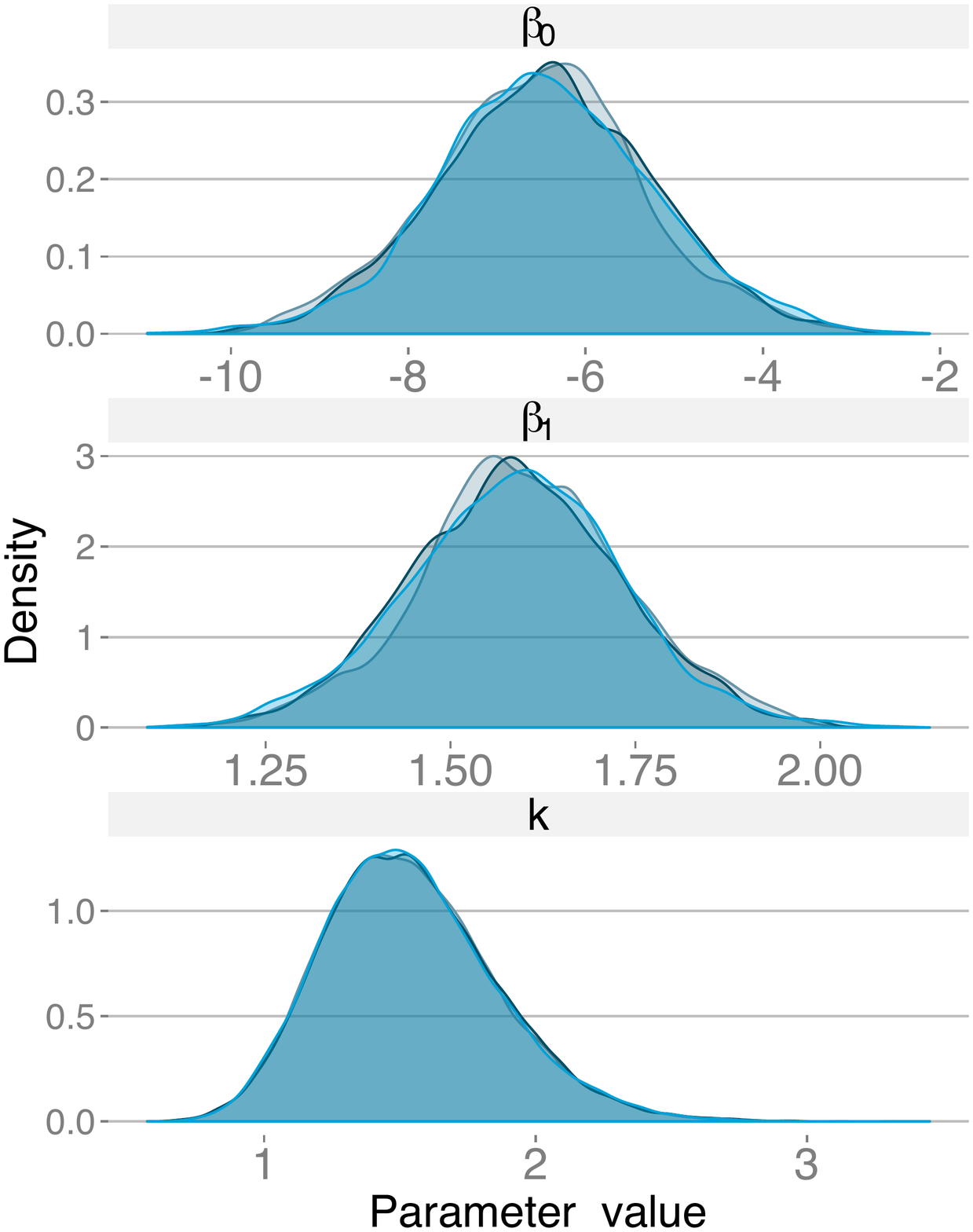}
\caption{Overlapped density plots with different colors by chain. The plot is a comparison of the target distribution by each chain, representing a visual test for convergence.  The figure displays the results for the model \ngc vs $M_{\rm BH}$, with the posteriors for  $\beta_0$, $\beta_1$ and $k$ displayed from top to bottom. 
}
\label{fig:density}
\end{figure}

If the Markov chains  are all representative of the posterior distribution of the fitted parameters, they should overlap each other.   Traceplots,  Fig.~\ref{fig:traceplot},  and density plots, Fig.~\ref{fig:density} are two useful visual diagnostics that are commonly used to test for chain convergence. We can see that the chains do mix well after the burn-in period, suggesting that the chains are producing representative values  from the posterior distribution for $\beta_0$, $\beta_1$ and $k$. Additionally, we used a more quantitative check, viz., the  so-called  Gelman-Rubin statistic \citep{gelman1992}. The underlying idea is that if the chains have reached convergence, the average difference between the chains should be similar to the average difference across steps within the chains.  The statistic equals unity if the chains are fully converged. As a rule, values above 1.1 indicate that the chains have failed to properly converge.   The Gelman-Rubin statistic fell below 1.05 for all estimated parameters in our analysis. Hence, once we convince ourselves that the model is working properly, the next step in the analysis is  to add  interpretations to the fitted coefficients as we discuss now.

\subsection{Interpretation of the  coefficients}

The  exponentiated coefficients $e^{\beta_i}$ of  Poisson and NB regressions are also  known as  rate ratios, or incidence rate ratios, which quantify how  an increase of unity in the predictor variable affects  the number of occurrences of the response variable.    
From Table~\ref{tab:betas}, displaying the means and respective 95\% credible intervals  of the  posterior distribution for each parameter,    the   exponentiated   coefficient  $\beta_1 = 1.59$ of  the \mbh predictor gives a  rate ratio of $e^{1.59}=4.9$. Therefore, according to the model, a  galaxy whose central black hole has a mass of, e.g., $\approx 10^8 M_{\bigodot}$   has on average approximately  five  times more globular clusters than a galaxy whose  $\mbh \approx 10^7 M_{\bigodot}$\footnote{Note that the analysis was made using  $\log{\mbh}$.}. In other words,  one dex\footnote{A dex difference of a given quantity x is a change by a factor of $10^x$.} 
variation increase in the \mbh leads to an approximately five  times increase in the incidence of globular clusters in a given  galaxy. Likewise, an increase of one dex in $M_{dyn}$ leads to an increase of $e^{2.19}=8.9$ times in  the population size of  globular clusters. Another way to state this is, given two galaxies with a difference in dynamical mass of one dex, the more massive one has a production rate of globular clusters 8.9 times more efficient  on average.  Similar interpretation can be made on the other parameters. 
Another question of interest is how  to determine the best predictor of \ngc. In the following, we discuss how to address this problem from a Bayesian perspective.

\begin{table}
\label{tab:coef}
\caption{$\beta_i$ coefficients and scale parameter,  $k$,   from Bayesian negative binomial  regression analysis with \ngc as the response variable and \mbh, $M_{\rm dyn}$, $\sigma$ and $M_V$  as  predictors. The upper and lower limits encloses 95\% of the  credible intervals around the posterior means.}
\label{tab:betas}
\begin{tabular}{l  c c  c }
\hline  
Predictor & $\beta_0$ & $\beta_1$ & k    \\
\hline\\[-1.5ex]
\mbh& $-6.49\pm 2.6$ & $1.59\pm 0.30$ & $1.53\pm 0.60$   \\[0.75ex]
$M_{\rm dyn}$ & $-17.72\pm 2.75$  &  $2.19\pm 0.24$ & $2.46\pm 0.97$   \\ [0.75ex]
$\sigma$ & $2.99\pm 0.78$  &  $0.02\pm 0.003$ & $1.52\pm 0.59$   \\ [0.75ex]
$M_{V}$ & $-20.50\pm 3.9$  &  $-1.28\pm 0.17$ & $2.23\pm 1.1$   \\ [0.75ex]
\hline 
\end{tabular}
\end{table}

\subsection{Model Comparison}

\begin{figure}
\centering
\includegraphics[width=1\columnwidth]{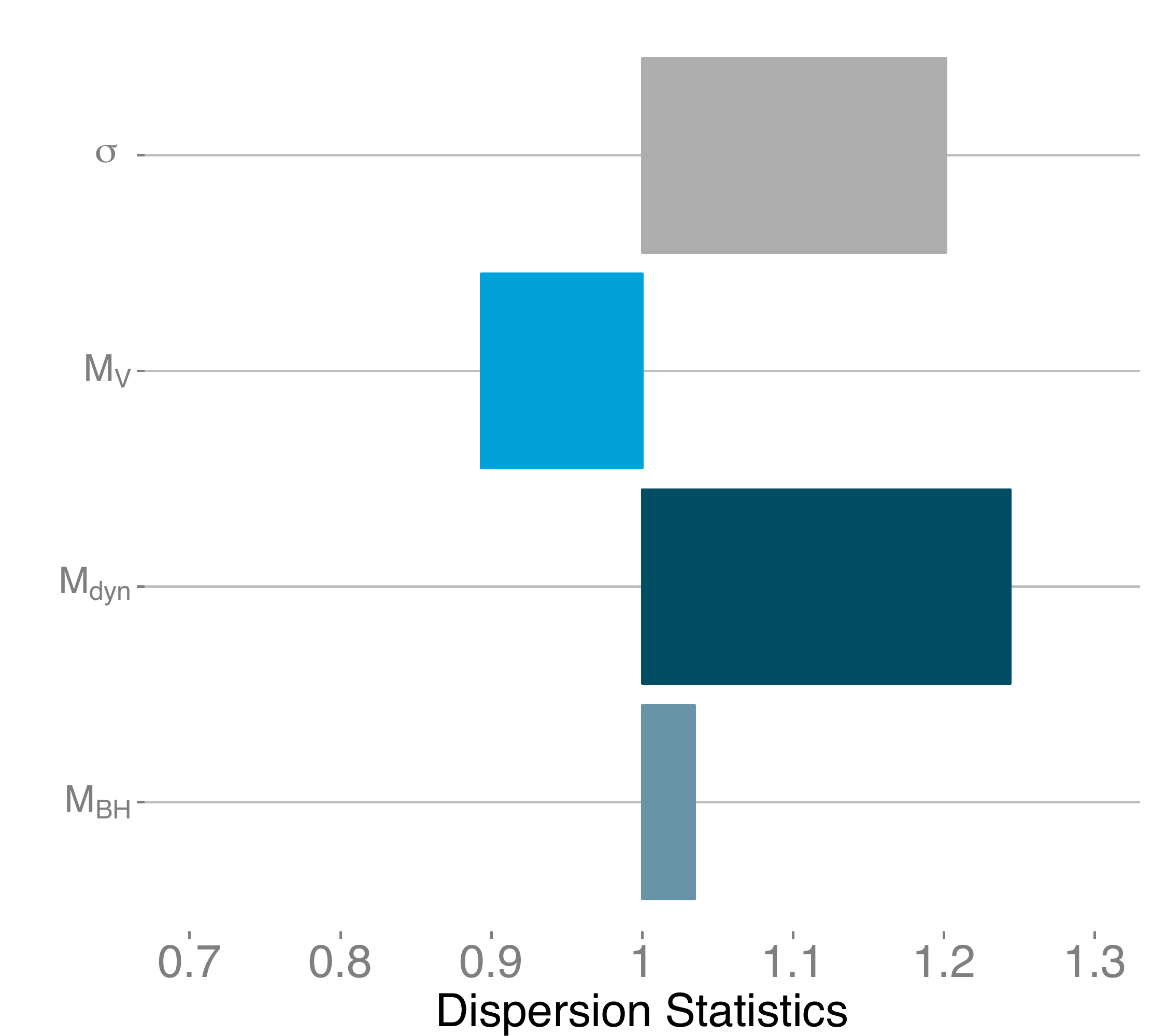}
\caption{Dispersion statistics,  $\mathcal{D}$, for each model. Values above 1 represent overdispersion, while values below 1 indicate underdispersion.  
}
\label{fig:Disp}
\end{figure}

\begin{figure}
\centering
\includegraphics[width=1\columnwidth]{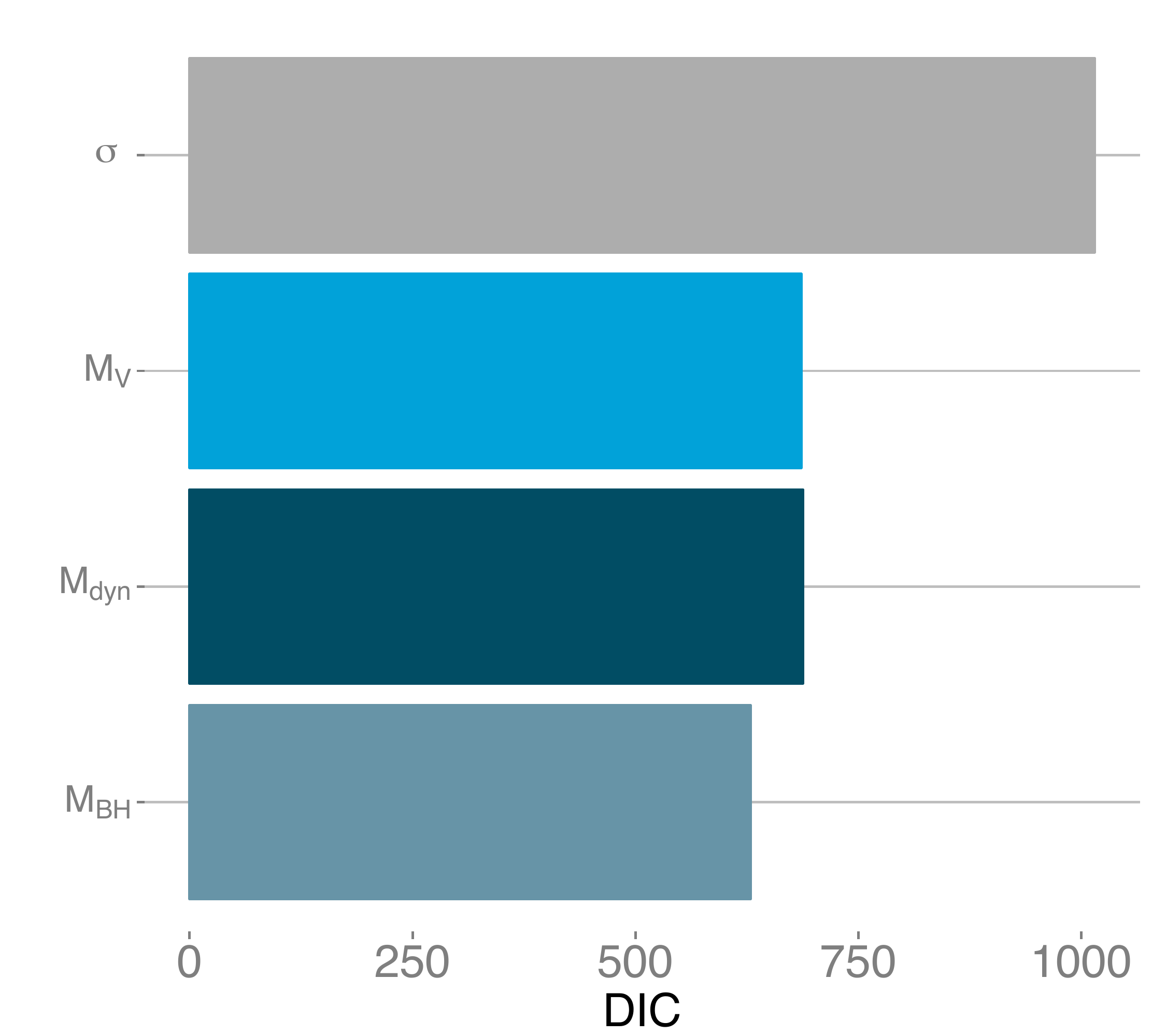}
\caption{Deviance information criterion, DIC, for each model. Smaller DIC values correspond to preferred  models. 
}
\label{fig:DIC}
\end{figure}

To find the best predictors for the globular cluster population, we compare the models using the dispersion statistics $\mathcal{D}$ defined in section~\ref{sec:overdispersion}, and the deviance information criterion \citep[DIC;][]{Spiegelhalter02}. The latter represents a compromise between the goodness of fit and model complexity. It is  defined as:
\begin{equation}
DIC = \overline{Dev}+p_D, 
\end{equation}
where the $\overline{Dev}$ is the average of the deviance $Dev(\mathbf{\theta})$ defined as   $Dev(\mathbf{\theta}) = -2\log\mathcal{L}(\rm{data}|\mathbf{\theta})$, with   $\mathcal{L}$ representing   the likelihood function. %
The effective number of parameters,  $p_D$, 
is calculated as:
\begin{equation}
p_D = \overline{Dev}-Dev(\mathbf{\theta}), 
\end{equation}
where   $\mathbf{\theta}$ is the vector of model parameters (${\beta_0,\beta_1, k}$ for the case in study  here). The preferred model has the smallest value for the DIC statistic.  Figs.~\ref{fig:Disp} and \ref{fig:DIC} depict the results for the model comparison using the same dataset. The black hole mass displays the lowest values for $\mathcal{D}$  and DIC, with dispersion statistics as low as  $\mathcal{D} = 1.05$. Although derived from an independent analysis, these findings corroborate   previous  claims about the tight connection between the central black hole mass and globular cluster population \citep{Burkert2010}. Nevertheless, it's worth noting that this is not in agreement with a previous analysis performed by  \citet{Harris2013}  using the same catalogue, where they  found $M_{\rm dyn}$ as a better  predictor for \ngc than $M_{\rm BH}$\footnote{It is important to note that we are not modelling the same relationship as \citeauthor{Harris2013} who modelled $\log N_{GC}$, the logarithm transformation of \ngc, while we model \ngc in the original scale.}.

%

\subsection{Further analysis with the entire data set}
\label{sec:all}

\begin{figure}
\centering
\includegraphics[width=1\columnwidth]{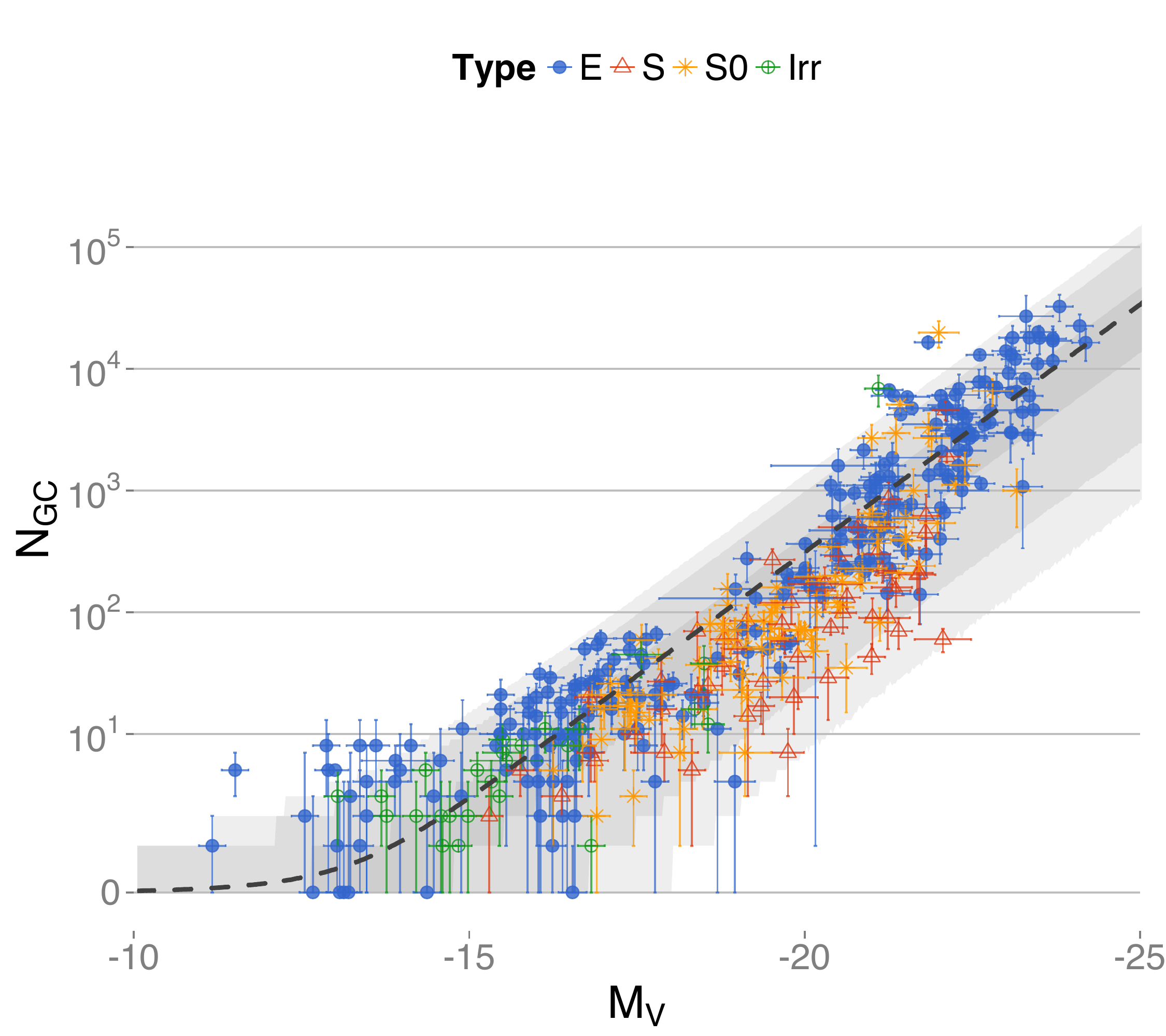}
\caption{Globular cluster population, $N_{\rm GC}$ plotted against visual absolute magnitude $M_V$. The dashed line   represents the expected value of \ngc for each value of $M_V$, while  
the shaded areas depicts 50\%,  95\%,  and  99\% prediction intervals. Galaxy types are coded by shape and colour as follows: Ellipticals (E; blue solid circles),  spirals (S; red open triangles), lenticulars (S0; orange asterisks), and irregulars (Irr; green open circles). An  ArcSinh transformation is applied in the y-axis for better visualization of the whole   range of $N_{\rm GC}$ values, including the null ones.
}
\label{fig:MV_full}
\end{figure}

\begin{figure}
\centering
\includegraphics[trim = 5mm 25mm 5mm 25mm, clip,width=1\columnwidth]{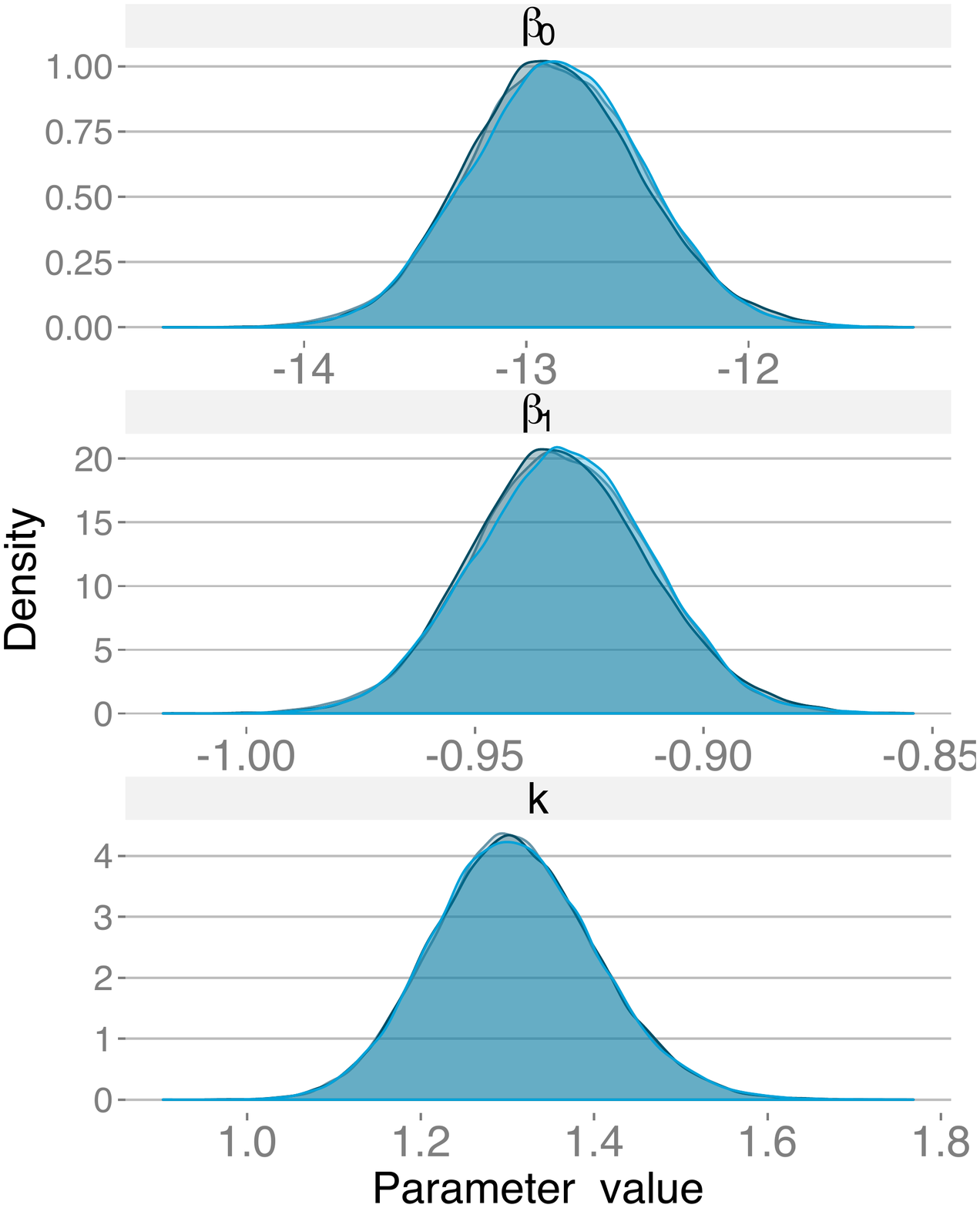}
\caption{Overlapped density plots with different colors by chain. The plot is a comparison of the target distribution by each chain, representing a visual test for convergence.  The figure displays the results for the model $N_{\rm GC}$ vs $M_{V}$, with the posteriors for  $\beta_0$, $\beta_1$ and $k$ displayed from top to bottom. 
}
\label{fig:density2}
\end{figure}

Hereafter, we provide a more extensive analysis using  the entire  catalogue of 422 galaxies. The only quantities available for all objects   are the \ngc, galaxy morphological type and $M_V$ \citep{Harris2013}. The advantage of using count models for this type of analysis is apparent from the six galaxies for  which no   globular clusters were detected. Such a scenario  is  naturally accommodated by   discrete likelihoods while avoiding the failings  of  logarithmic transformations to the response. The statistical model we use is the same as that discussed in the beginning of this section and can be described as:
\begin{alignteo}
\label{eq:HBMMV}
  &N_{\mathrm{GC};i}\sim \rm NB(p_i,k); \notag   \\
  &p_i =  \frac{k}{k+\mu_i}; \notag \\
   &\mu_i = e^{\eta_i} + \epsilon_{N_{GC};i}; \notag  \\ 
  &\eta_i =  \beta_0+\beta_1\times M_{V;i}^*; \notag \\
  & k \sim \mathcal{U}(0,5); \notag \\
  &M_{V;i} \sim \mathcal{N}(M_{V;i}^*,e_{M_{V};i}^2);\notag \\ 
  &\epsilon_{N_{GC};i} \sim \mathcal{B}(0.5, 2 e_{N_{GC};i}) - e_{N_{GC};i}; \\  
  &\beta_0 \sim \mathcal{N}(0,10^{6});\notag \\
  &\beta_1 \sim \mathcal{N}(0,10^{6});\notag \\
  &M_{V;i}^* \sim \mathcal{U}(-26,-10); \notag \\
  &i = 1,\cdots,N.\notag
\end{alignteo}
Overall, the model is similar to the one described in equation~(\ref{eq:HBM}). The difference is in the prior for the unobserved true absolute visual magnitude,  $M_{V;i}^*$,  to which we assigned a uniform prior over the range of magnitudes covered by the catalogue. 
The fitted model shows remarkable agreement with the data as displayed in Fig.~\ref{fig:MV_full}.  Very few objects fall outside the prediction intervals over  a wide range of galaxy brightnesses. 
The dispersion statistics for this model  is  $\mathcal{D} = 1.15$, and the credible intervals for the fitted $\beta$ coefficients and scaling parameter, $k$,  are shown in Fig.~\ref{fig:density2}.  Likewise, as in the previous section, we can  interpret the $\beta$ coefficient  as follows. 
The  mean value of $\beta_1$ exponentiated is  $\approx 0.4$, which implies that a galaxy whose  absolute visual   magnitude is one unit greater than another reference galaxy  has on average 0.4 times less globular clusters; i.e., a galaxy  brighter by one magnitude over another has on  average 2.5 times more globular clusters. 
Likewise, a galaxy with $M_V = -20$ has on average $2.5^5 \approx 100$ times more globular clusters than a galaxy with  $M_V = -15$, which is consistent with a visual inspection of Fig.~\ref{fig:MV_full}. Another advantage of our approach is the possibility to extrapolate the regression solution without making non-physical   predictions.  The fitted model  predicts a nearly zero occurrence of globular clusters for galaxies with $M_V \geqslant -11$. Considering  the  total galaxy  luminosity,  $L = 10^{0.4(M_{V_{\bigodot}}-M_V)}L_{\bigodot}$, with $M_{V_{\bigodot}} = 4.83$, the model suggests that galaxies with   $L \leqslant 2\times 10^6~L_{\bigodot}$  are unlikely to host populations of globular clusters, thus agreeing with the literature \citep[e.g.,][]{Harris2013}. The use of Bayesian prediction intervals allow us to make some interesting predictions: for instance  from  Fig.~\ref{fig:MV_full}, we can state that  galaxies with luminosities $L \leqslant 8.5\times 10^7~L_{\bigodot}$ (or $M_{V_{\bigodot}} \geqslant -15$)  should not contain  more than 10 globular clusters with 99\% probability.

The analysis performed so far did not account for  information regarding different  galaxy morphological types. Therefore, we are   implicitly assuming a pooled estimate \citep[e.g.,][]{Gelman2007}: all different galaxy types  are sampled from the same common distribution   ignoring any possible  variation among them.  On the other extreme, performing an independent analysis for each class would mean making the assumption that  each morphological type is sampled from independent  distributions and that  variations  between them cannot be combined.   In the next section we discuss a more flexible approach  together with a brief  overview of generalized linear mixed models. 

\section{Generalized Linear Mixed Models}
\label{sec:GLMMs}
As our final analysis, we introduce one of the most important extensions of the  GLM methodology  known as generalized linear mixed models (GLMMs). In particular, we focus on one  of the simplest GLMM  incarnations known as the random intercepts model. The random intercepts model, in our context, includes an additional term $\zeta_j$ to account for a class (galaxy type) specific deviation from the common  intercept $\beta_0$:
\begin{equation}
\eta_{ij} = \beta_0 + \beta_1 \times M_{V;i} + \zeta_j, 
\label{eq:intercept}
\end{equation}
where the index  $j$ runs from 1 to 69  representing each of the different galaxy subtypes reported in \citeauthor{Harris2013}. A standard approach  to modelling $\zeta_j$ in a standard  linear mixed regression model is to  assume  the conditional normality of the random intercepts with  $\zeta_j \sim \mathcal{N}(0,1/\tau)$, and  $\tau \sim \Gamma(0.01,0.01)$. Our intention in  incorporating this extra term into the model is not to simply adjust the data, but rather the aim is to identify any particular galaxy sub-type which deviates from the overall population mean. For this  purpose, we employed a popular method for variable selection from a Bayesian perspective known as least absolute shrinkage and
selection operator (LASSO) which is discussed in the following section.

\subsection{Bayesian LASSO}

\begin{figure}
\centering
\includegraphics[width=1\columnwidth]{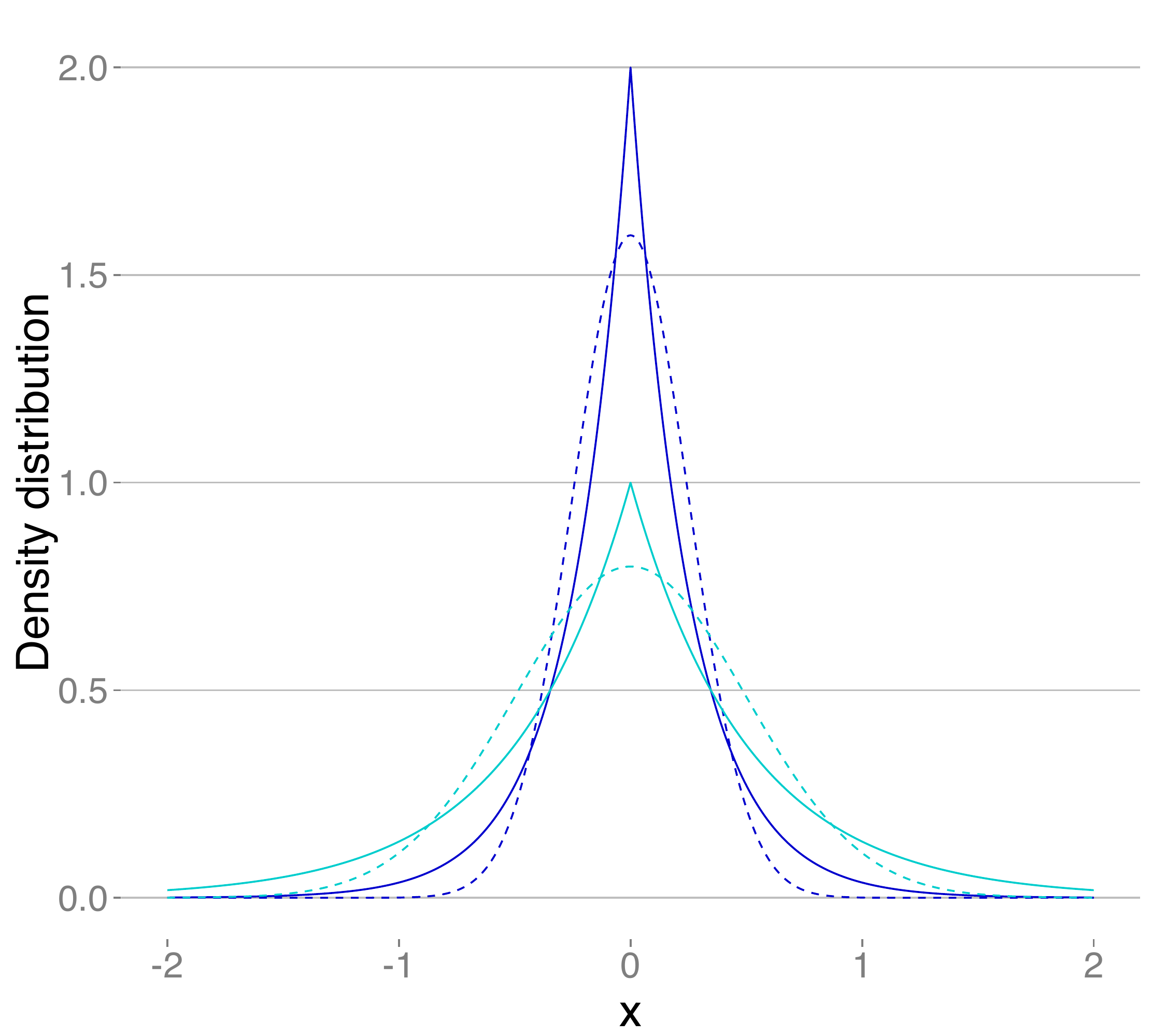}
\caption{Illustrative comparison between Laplace and Gaussian  priors. The Gaussian distribution is represented by dashed lines, while the Laplace distribution by solid lines. For all curves we assign a zero mean, and the scale (or standard deviation, $\sigma$,  for the  Gaussian case) parameters  0.25 (dark blue lines) and 0.5 (cyan lines).
}
\label{fig:priors}
\end{figure}

The original LASSO regression was proposed by  \citet{Tibshirani96} to automatically select a relevant subset of predictors in a regression problem by shrinking some 
coefficients towards zero \citep[see also][for a recent application of  LASSO for modelling  Type Ia supernovae light curves]{Uemura2015}.  For a  typical linear regression problem:
\begin{equation}
y_i=\beta_0+\beta_1x_1+\cdots+\beta_{p}x_p+\epsilon,
\end{equation}
with $\epsilon$ denoting Gaussian noise, LASSO estimates linear regression coefficients $\mathbf{\beta} = \beta_0+\beta_1x_1+\cdots+\beta_{p}x_p$  
by imposing a $L_1$-norm penalty in the form:
\begin{equation}
\argmin_{\beta} \left \{\sum_{i=1}^N\left(y_i-\sum_{j=1}^p\beta_jx_{ij}\right)^2+\kappa\sum_{j=1}^p|\beta_j| \right \}, 
\end{equation}
where $\kappa\geq 0$ is a tunable constant that controls the level of sparseness of the solution. The number of zero coefficients thereby increases as $\kappa$ increase. \citeauthor{Tibshirani96} also noted that the LASSO estimate has a Bayesian counterpart when the $\mathbf{\beta}$ coefficients have a double-exponential prior (i.e., a Laplace prior) distribution,
\begin{equation}
f(\mathbf{\zeta};\tau) = \frac{1}{2\tau}\exp\left(-\frac{|\zeta_j|}{\tau}\right), 
\end{equation}
where $\tau = 1/\kappa$. The idea was further developed and is known as Bayesian LASSO \citep[see e.g., ][]{Park2008}.
Hereafter,  we use the LASSO formulation  for a slightly  different purpose, viz., variable selection for random intercept models \citep[see e.g.,][pg. 165]{bernardo2011}.  
The underlying idea is to discriminate between  galaxy types that follow the overall population mean, i.e. $\zeta_1 = 0$, and galaxies that require an additional adjustment in the intercept, i.e. $\zeta_i \neq 0$.
In order to include this information, we replace the linear predictor $\eta$ by equation~(\ref{eq:intercept}) and add the following equations in the model described by equation~(\ref{eq:HBMMV}):
\begin{alignteo}
\label{eq:lasso}
  &\zeta_j \sim Laplace\left(0,\tau\right); \notag \\
  &\tau = 1/\kappa;  \\
   &\kappa \sim  \Gamma(0.01,0.01); \notag \\
  &j = 1,\cdots,69.\notag
\end{alignteo}
The role of the Laplace prior  is to  assign more weight to regions either  near to  zero or in the distribution  tails as  compared to  a normal prior. A visual inspection on   Fig.~\ref{fig:priors} confirms this notion. For the parameter $\kappa$, we assigned a  diffuse (non-informative) gamma hyperprior  in the form $\kappa \sim \Gamma(0.01,0.01)$, which avoids the need of an ad hoc choice of $\kappa$.  Note that  other possibilities exist such as, e.g.,  iteratively finding $\kappa$ via cross-validation to maximize predictive power. 

\begin{figure}
\centering
\includegraphics[width=1\columnwidth]{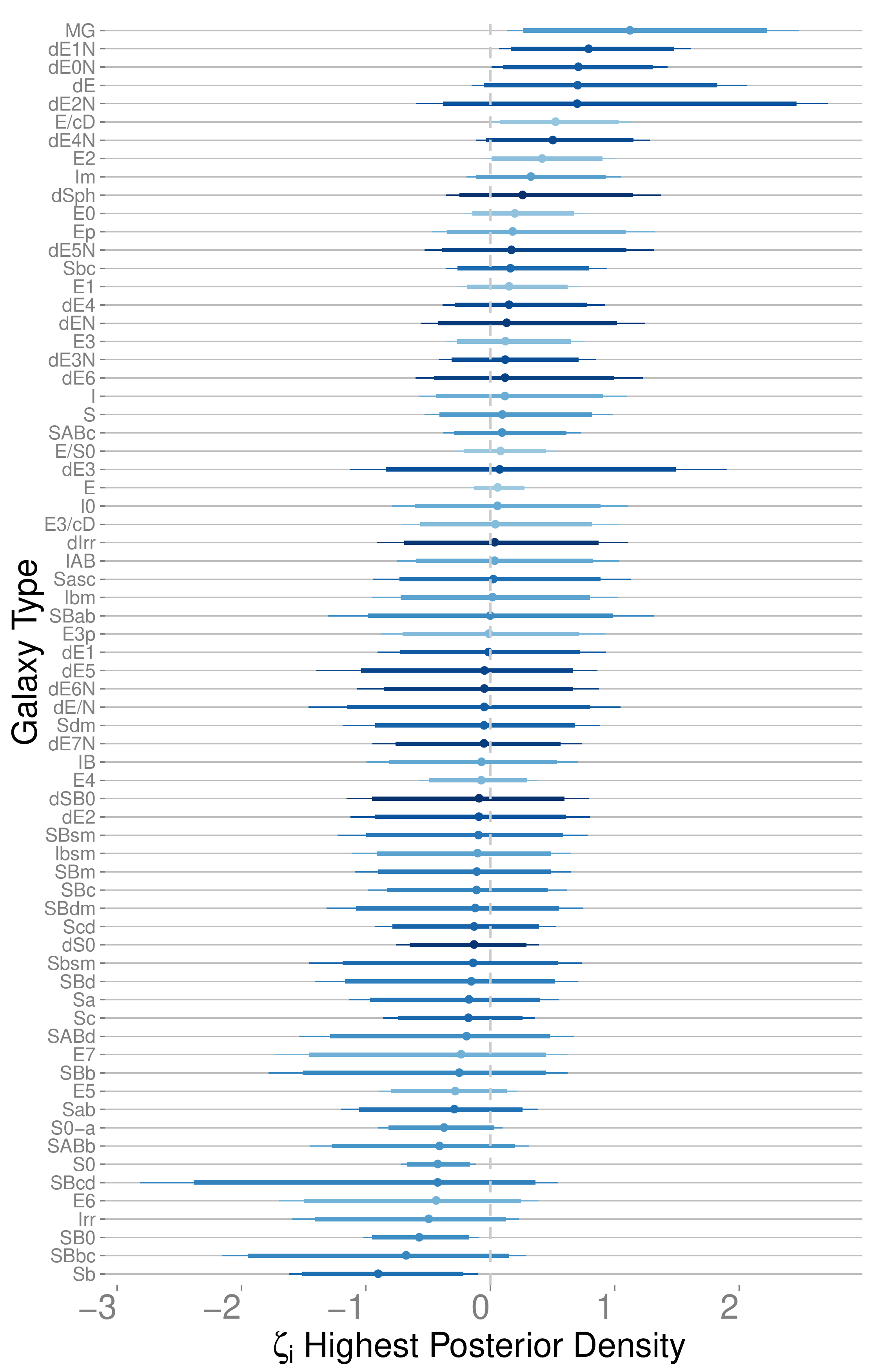}
\caption{Caterpillar plot for the random intercepts $\zeta_i$ versus the subcategories of galaxy morphological classifications. The thick and thin horizontal lines represent 90\% and 95\% credible intervals respectively. 
}
\label{fig:random}
\end{figure}

\begin{figure}
\centering
\includegraphics[width=1\columnwidth]{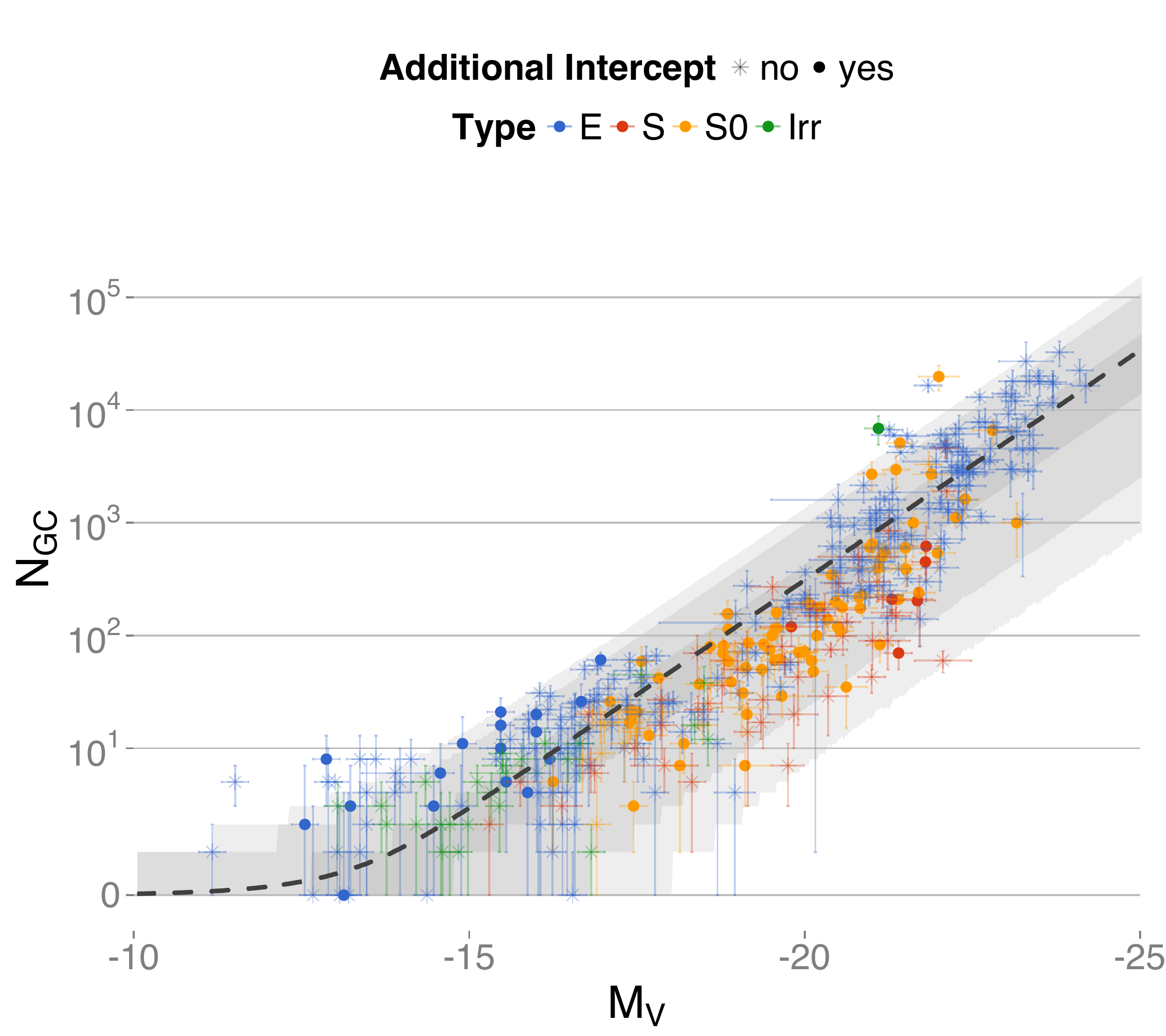}
\caption{Globular cluster population, \ngc plotted against visual absolute magnitude $M_V$. The dashed line   represents the expected value of \ngc for each value of $M_V$, while  
the shaded areas depicts 50\%,  95\%,  and  99\% prediction intervals. Galaxy types are coded by  colours as follows: Ellipticals (E; blue),  spirals (S; red), lenticulars (S0; orange), and irregulars (Irr; green). Asterisks represent galaxies belonging to sub-types whose random intercept $\zeta$ is consistent with zero, while circles represent the  ones with $\zeta \neq 0$. 
An  ArcSinh transformation is applied in the y-axis for better visualization of the whole   range of \ngc values, including the null ones.
}
\label{fig:MV_fullout}
\end{figure}

Analysis results are displayed on Fig.~\ref{fig:random}. Overall, it suggests that we do not need to add an additional intercept  for predicting \ngc  from  $M_V$. This is consistent with the fact that prediction intervals in Fig.~\ref{fig:MV_full} enclose $\sim 98.8\%$  of the data set without any need of a random intercept. Nevertheless, the following  galaxy types require systematic adjustments:  spirals galaxies with moderate size of nuclear bulge
(Sb), barred lenticulars  (SB0), lenticulars (S0) and  dwarf  elliptical galaxies (dE0N and dE1N). MG represents one single  
object.  Also, UGC 3274 is the brightest galaxy of the galaxy cluster ACO 539 \citep{Lin2004}. Fig.~\ref{fig:MV_fullout}  shows that the dE0N and dE1N objects  have a large number of GCs on average when compared to other galaxy types with similar luminosities, while the lenticulars have systematically fewer GCs  
than expected for the overall galaxy population. This can be  quantified by looking at the mean value of $\zeta$ in Fig.~\ref{fig:random}. For S0 galaxies the mean value of  $\zeta$ is   -0.42 indicating that, on average, S0 galaxies have 34\% ($1-e^{-0.42}$) fewer GCs than other galaxy types in the same range of luminosities. Generally speaking, galaxy  types with  95\% credible intervals falling on the right side of the dashed grey  vertical line in Fig.~\ref{fig:random} have more GCs than the overall population mean, while galaxy types on the left side have fewer GCs than the population mean.  While a detailed investigation of the causes of this behaviour is beyond the scope of this work, it is important to stress the ability  of  hierarchical Bayesian models to explore the multilevel statistical properties of the objects under study in an unified way.

\section{Conclusions}
\label{sec:end}

We employed a Bayesian negative binomial  regression model to analyse the population size of globular clusters in the presence of galactic attributes such as central black hole mass, brightness, and morphological type. Hence,  demonstrating  how  generalized linear models designed to represent count data 
provide reliable outcomes and interpretations. The main scientific   results and features of our analysis can be summarized as follows:

\begin{itemize}

\item The population size of GC is on average 35\% lower on S0 galaxies if compared to other galaxies with similar luminosities.

\item The relationship between the number of globular clusters and other galaxy properties has more variation than expected by a Poisson process, but can be  well modelled by a negative binomial GLM. 

\item The Bayesian modelling herein employed naturally accounts for  heteroscedasticity, intrinsic scatter, and  errors in measurements in both axes (either discrete or continuous).

\item Predicted intervals around the trend for expected \ngc envelope the data, including the Milky Way, which was previously considered an outlier.

\item The random intercepts model (with a Bayesian LASSO)  applied to the correlation between GC population and  brightness allow us to account for the presence of 69 different galaxy subcategories of morphological classifications, and automatically identifies particular types not following the overall population mean. Galaxy types dE1N, dE0N, E/cD, S0, Sb0 and Sb show significant deviations from the general trend. Based on the sample studied here, we advise these types to be further scrutinized in order to clarify if there is any physical mechanism behind  such deviations or  merely an observational bias\footnote{Type MG also shows significant deviation, but this is probably a consequence of small sample size (MG corresponds to only 1 object).}. 

\item By employing a hierarchical Bayesian model for the  random intercepts and unobserved covariates (e.g., true black hole mass), we allow the model to  borrow  strength across units. This happens via their joint influence on
the posterior estimates of the unknown hyper-parameters.

\item If extrapolated, the fitted model predicts  a  suppression in the presence of GCs for galaxies with luminosities  $L \lesssim 2\times 10^6~L_{\bigodot}$. 

\item The central black holes mass is in fact a good predictor of the number of GCs. One dex increase in $M_{BH}$ leads to an approximate 5 times increase in the incidence of globular clusters. The origin of such correlation  it is still a matter of debate. One possible explanation is that both properties are associated with a common event such as major mergers, thus galaxies experimenting  a recent major merger should  have a  large $M_{BH}$  mass and GC populations \citep[e.g.,][]{Jahnke2011}.  
The total mass of GCs and the central black hole  mass can also  correlate 
with the  bulge binding energy in elliptical galaxies  \citep[e.g.,][]{Snyder2011,Saxton2014}.  Rapid growth of the nuclear black hole of a galaxy  might be fuelled by a massive inflow of cold gas towards the centre of the galaxy. The gas inflow would trigger star formation  and the formation of  GCs. Hence, leading to an indirect  correlation between the  total number of GCs and the $M_{BH}$.
Scrutinizing which one among these and other possibilities, if any, are responsible for this correlation (causal or not) is beyond the purposes of this work. However, it does provide a clear example on how the adoption of modern statistical methods can point to intriguing astrophysical questions.

\end{itemize}

A statistical model is based on an appropriate probability distribution function assumed to generate or describe a data set. Hence, the parameter estimating likelihood function must specify a probability distribution on the appropriate scale under study.  
Discrete data, and count data in particular, are not continuous as are data described by the Gaussian distribution. The most appropriate way to model count data is by using a discrete probability distribution,  e.g., a Poisson or negative binomial likelihood, otherwise the model will likely be biased and misspecified --- the price to be paid for employing the wrong likelihood estimator for the data of interest. 

Generalized linear models are a  cornerstone of modern statistics, and an invaluable instrument for astronomical investigations given their potential  application to a variety of astronomical problems beyond Gaussian assumptions. A prompt integration of these methods into astronomical analyses will allow  contemporary statistical techniques  to become common practice in the research of $\rm 21^{st}$ century astronomy.

\section*{Acknowledgements}
We thank Johannes Buchner for careful revision and  constructive comments. 
 The IAA Cosmostatistics Initiative (COIN)\footnote{\url{https://asaip.psu.edu/organizations/iaa/iaa-working-group-of-cosmostatistics}} is a nonprofit organization whose aim is to nourish the synergy between astrophysics, cosmology, statistics and machine learning communities.
EEOI is partially supported by the Brazilian agency CAPES (grant number 9229-13-2). ACS acknowledges funding from a CNPq, BJT-A fellowship (400857/2014-6). MK acknowledges support by the DFG project DO 1310/4-1. This work was written on the collaborative \texttt{Overleaf} platform\footnote{\url{www.overleaf.com}}, and made use of the GitHub\footnote{\url{www.github.com}} repository web-based hosting service and \texttt{git} version control software. Work on this paper has substantially benefited from using the collaborative website AWOB (http://awob.mpg.de) developed and maintained by the Max-Planck Institute for Astrophysics and the Max-Planck Digital Library.
The bibliographic research was possible thanks to the tools offered by the NASA Astrophysical Data Systems.


\appendix

\section{JAGS model}
\label{app:JAGS}

\lstset{
language=R,
keywordstyle=\bfseries\ttfamily\color[rgb]{0,0,1},
identifierstyle=\ttfamily,
commentstyle=\color{MidnightBlue},
stringstyle=\ttfamily\color{blue},
numbersep=5pt,
xleftmargin=20pt,
frame=tb,
framexleftmargin=20pt,
showstringspaces=false,
	basicstyle=\footnotesize,
	numberstyle=\tiny,
	numbers=left,
	stepnumber=1,
	numbersep=10pt,
	tabsize=2,
	breaklines=true,
	prebreak = \raisebox{0ex}[0ex][0ex]{\ensuremath{\hookleftarrow}},
	breakatwhitespace=false,
	aboveskip={1.5\baselineskip},
  columns=fixed,
  extendedchars=true
}

\paragraph*{Poisson GLM}
The basic JAGS syntax  for a Poisson GLM model:

\begin{lstlisting}
GLM.pois<-model{
#Priors for regression coefficients

beta.0~dnorm(0,0.000001)
beta.1~dnorm(0,0.000001)

#Poisson GLM Likelihood

for (i in 1:N){
eta[i]<-beta.0+beta.1*x[i]
log(mu[i])<-eta[i]
y[i]~dpois(mu[i])
              }
}

\end{lstlisting} 

\paragraph*{Negative Binomial GLM}
The basic JAGS syntax  for a NB GLM model:

\begin{lstlisting}
GLM.NB<-model{
#Priors for regression coefficients

beta.0~dnorm(0,0.000001)
beta.1~dnorm(0,0.000001)
k~dunif(0.001,10)

#NB GLM Likelihood

for (i in 1:N){
eta[i]<-beta.0+beta.1*x[i]
log(mu[i])<-eta[i]
p[i]<-k/(k+mu[i])
y[i]~dnegbin(p[i],k)
              }
}

\end{lstlisting} 
Another approach to fit a NB model in JAGS is via a combination of a Gamma distribution with a Poisson distribution in the form \citep[see e.g.,][]{Marley2010,Hilbe2011}:

\begin{lstlisting}
GLM.NB<-model{
#Priors for regression coefficients

beta.0~dnorm(0,0.000001)
beta.1~dnorm(0,0.000001)
k~dunif(0.001,10)

#NB GLM Likelihood

for (i in 1:N){
eta[i]<-beta.0+beta.1*x[i]
log(mu[i])<-eta[i]
rateParm[i]<-k/mu[i]
g[i]~dgamma(k,rateParm[i])
y[i]~dpois(g[i],k)
              }
}

\end{lstlisting} 

\section{Bayesian model for each covariate}
\label{app:hbm}


\paragraph*{Dynamical mass versus globular cluster population}

Bayesian NB GLM model for the relationship between \ngc and galaxy dynamical mass $M_{dyn}$. Since,  there is no information about the uncertainties in the measurements of $M_{dyn}$, we neglect this in this particular model.  

\begin{alignteo}
\label{eq:HBMMdyn}
  &N_{GC;i}\sim \rm NB(p_i,k); \notag   \\
  &p_i =  \frac{k}{k+\mu_i}; \notag \\
   &\mu_i = e^{\eta_i} + \epsilon_{N_{GC};i}; \notag  \\
  &\eta_i =  \beta_0+\beta_1\times M_{dyn;i}; \notag \\
  & k \sim \mathcal{U}(0,5); \notag \\
  &\epsilon_{N_{GC};i} \sim \mathcal{B}(0.5, 2 e_{N_{GC};i}) - e_{N_{GC};i}; \\  
  &\beta_0 \sim \mathcal{N}(0,10^{6});\notag \\
  &\beta_1 \sim \mathcal{N}(0,10^{6});\notag \\
  &\alpha_0 \sim \Gamma(0.01,0.01); \notag \\
  &\theta_0 \sim \Gamma(0.01,0.01); \notag \\
  &i = 1,\cdots,N.\notag
\end{alignteo}

\paragraph*{Bulge velocity versus globular cluster population}
Bayesian NB GLM model for the relationship between \ngc and bulge dispersion velocity $\sigma$. 

\begin{alignteo}
\label{eq:HBMsigma}
  &N_{GC;i}\sim \rm NB(p_i,k); \notag   \\
  &p_i =  \frac{k}{k+\mu_i}; \notag \\
   &\mu_i = e^{\eta_i} + \epsilon_{N_{GC};i}; \notag  \\
  &\eta_i =  \beta_0+\beta_1\times \sigma_i^*; \notag \\
  & k \sim \mathcal{U}(0,5); \notag \\
  &\sigma_i \sim \mathcal{N}(\sigma_i^*,e_{\sigma_i}^2);\notag \\ 
  &\epsilon_{N_{GC};i} \sim \mathcal{B}(0.5, 2 e_{N_{GC};i}) - e_{N_{GC};i}; \\  
  &\beta_0 \sim \mathcal{N}(0,10^{6});\notag \\
  &\beta_1 \sim \mathcal{N}(0,10^{6});\notag \\
  &\sigma_i^* \sim \Gamma(\alpha_0,\theta_0); \notag \\
  &\alpha_0 \sim \Gamma(0.01,0.01); \notag \\
  &\theta_0 \sim \Gamma(0.01,0.01); \notag \\
  &i = 1,\cdots,N.\notag
\end{alignteo}

\end{document}